\documentclass[10pt,draftclsnofoot,onecolumn]{IEEEtran}
\usepackage{graphicx} %you can incorporate graphics in pdf, jpg, png or tif(with one f) formats. If you don't specify the file extension, it will look for and use whichever one is available.
\IEEEoverridecommandlockouts
\usepackage[mode=buildnew]{standalone}
\usepackage{comment}
\graphicspath{{Figures/}}
\usepackage[sort&compress, numbers]{natbib}
\usepackage{amsmath}
\usepackage{array}
\usepackage{url}
\usepackage{amssymb}
\usepackage{stmaryrd}
\SetSymbolFont{stmry}{bold}{U}{stmry}{m}{n}
\usepackage{mathtools}
\usepackage{amsthm}
\usepackage{tabu}
\usepackage{xcolor}
\usepackage{colortbl}
\usepackage{fullpage}
\usepackage[caption=false,font=footnotesize]{subfig}
\usepackage{multirow}
\usepackage{soul}
\usepackage{tikz}
\usetikzlibrary{arrows}

\usepackage{pgfplots}
\pgfplotsset{compat=1.18}
\usepackage{color}
\usepackage{xr-hyper}
\usepackage{hyperref}
\makeatletter
\newcommand*{\addFileDependency}[1]{% argument=file name and extension
  \typeout{(#1)}
  \@addtofilelist{#1}
  \IfFileExists{#1}{}{\typeout{No file #1.}}
}
\makeatother

\usepackage[ruled,linesnumbered]{algorithm2e}

\SetCommentSty{mycommfont}

\SetKwInput{KwInput}{Input}                % Set the Input
\SetKwInput{KwOutput}{Output}              % set the Output

\DeclareMathOperator*{\argmin}{arg\,min}

\newcommand{\bs}{\boldsymbol}
\newcommand{\bb}{\mathbb}
\newcommand{\cl}{\mathcal}

\newcommand{\iid}{% independent and identically distributed
    \ifmmode% math mode
        \mathrm{iid}%
    \else%
        iid\xspace%
    \fi%
}
\newcommand{\ie}{\emph{i.e.}, }
\newcommand{\eg}{\emph{e.g.}, }
\newcommand{\etal}{\emph{et al.}}

\newcommand{\sqp}{\vspace{2mm}}

\newcommand{\dbracket}[1]{\llbracket #1 \rrbracket}
\newcommand{\algname}{LoRePIE }
\newcommand{\Hp}{H_{\rm p}}
\newcommand{\Wp}{W_{\rm p}}
\newcommand{\Np}{N_{\rm p}}

\newcommand{\Hd}{H_{\rm d}}
\newcommand{\Wd}{W_{\rm d}}
\newcommand{\Nd}{N_{\rm d}}
\newcommand{\No}{N_{\rm o}}
\newcommand{\Ho}{H_{\rm o}}
\newcommand{\Wo}{W_{\rm o}}
\renewcommand{\No}{N_{\rm o}}
\begin{document}
\title{LoRePIE: $\ell_0$ Regularised \\ Extended Ptychographical Iterative Engine \\ for Low-dose and Fast Electron Ptychography
\thanks{Alex. W. Robinson is now affiliated with “SenseAI 
Innovations Ltd., Brodie Tower, University of Liverpool, Liverpool, 
UK.” Nigel D. Browning is now affiliated with both “Department of 
Mechanical, Materials, \& Aerospace Engineering, University of Liverpool, Liverpool, UK” and “SenseAI Innovations Ltd., Brodie Tower, 
University of Liverpool, Liverpool, UK.”.}}
\author{\IEEEauthorblockN{
    Amirafshar~Moshtaghpour\IEEEauthorrefmark{1}\IEEEauthorrefmark{2}, 
    Abner~Velazco-Torrejon\IEEEauthorrefmark{1},
    Alex~W.~Robinson\IEEEauthorrefmark{2},\\
    Nigel~D.~Browning\IEEEauthorrefmark{2}\IEEEauthorrefmark{3}, and Angus~I.~Kirkland\IEEEauthorrefmark{1}\IEEEauthorrefmark{4}
    }
    \\
    \IEEEauthorblockA{
        \IEEEauthorrefmark{1} Rosalind Franklin Institute, Harwell Science \& Innovation Campus, Didcot, OX11 0QS, UK.\\
        \IEEEauthorrefmark{2} Mechanical, Materials, and Aerospace Engineering, University of Liverpool, Liverpool, L69 3GH, UK.\\
        \IEEEauthorrefmark{4} Department of Materials, University of Oxford, Oxford, OX2 6NN, UK.}
}
\maketitle
\begin{abstract}
The extended Ptychographical Iterative Engine (ePIE) is a widely used phase retrieval algorithm for Electron Ptychography from 4-dimensional (4-D) Scanning Transmission Electron Microscopy (4-D STEM) measurements acquired with a focused or defocused electron probe. However, ePIE relies on redundancy in the data and hence requires adjacent illuminated areas to overlap. In this paper, we propose a regularised variant of ePIE that is more robust to low overlap ratios. We examine the performance of the proposed algorithm on an experimental 4-D STEM data of double layered Rotavirus particles acquired in a full scan with 85\% overlap. By artificial down-sampling of the probe positions, we have created synthetic 4-D STEM datasets with different overlap ratios and use these to show that a high quality reconstruction of Rotavirus particles can be obtained from data with an overlap as low as 56\%.
\end{abstract}
\section{Introduction}\label{sec:introduction}
The Scanning Transmission Electron Microscope (STEM) with a coherent electron source and a spherical aberration corrector is the state-of-the-art tool for studying complex structures at atomic resolution~\cite{krivanek1999towards,batson2002sub}. Importantly, multiple signals can be collected simultaneously to retrieve structural and analytical information. Compositional contrast can be accessed through electrons scattered at high angles \cite{HAADFHowie1979,HAADFPennycook1989}, while electronic properties can be studied using electron energy loss spectroscopy (EELS). This capability makes it an attractive tool in, \eg material science \cite{PENNYCOOK201722,chen2021electron}, biological sciences \cite{Bio_STEM2012,zhou2020lowdose} and the semiconductor industry~\cite{van2009imaging,chejarla2023measuring}. 

Despite this potential, the limiting factor in STEM is the sensitivity of the material under study to the electron beam. Light elements are more sensitive to the electron beam and exhibit lower contrast compared to heavy elements~\cite{EGERTON2004399}. In addition, insulating materials are subject to increased radiolytic damage compared to electrical conductors~\cite{EGERTON2004399}. A common strategy to mitigate electron beam damage is to reduce the electron fluence. However, this approach results in images with a reduced signal-to-noise ratio (SNR), which makes the study of many classes of materials difficult. Therefore, alternative methods that make use of a limiting electron fluence budget in more efficient ways have been developed. Examples of such methods, based on the dynamic nature of beam damage, utilise a non-conventional interleaved scan~\cite{velazcoreducing2022} or are based on subsampling the probe positions~\cite{nicholls2022compressive}. Advances in pixelated direct electron detectors \cite{pix_detectorsI, pix_detectorsII} have further enabled 4-dimensional (4-D) STEM. Among the different signals of interest that can be obtained from a 4-D STEM dataset, phase contrast imaging is an attractive alternative for imaging low-contrast soft materials under low-fluence conditions ~\cite{Ophus_2019,song2019atomic,zhou2020lowdose}.

\begin{figure}[!t]
    \centering
\includegraphics[width=0.42\columnwidth,trim={0 5mm 0 0},clip]{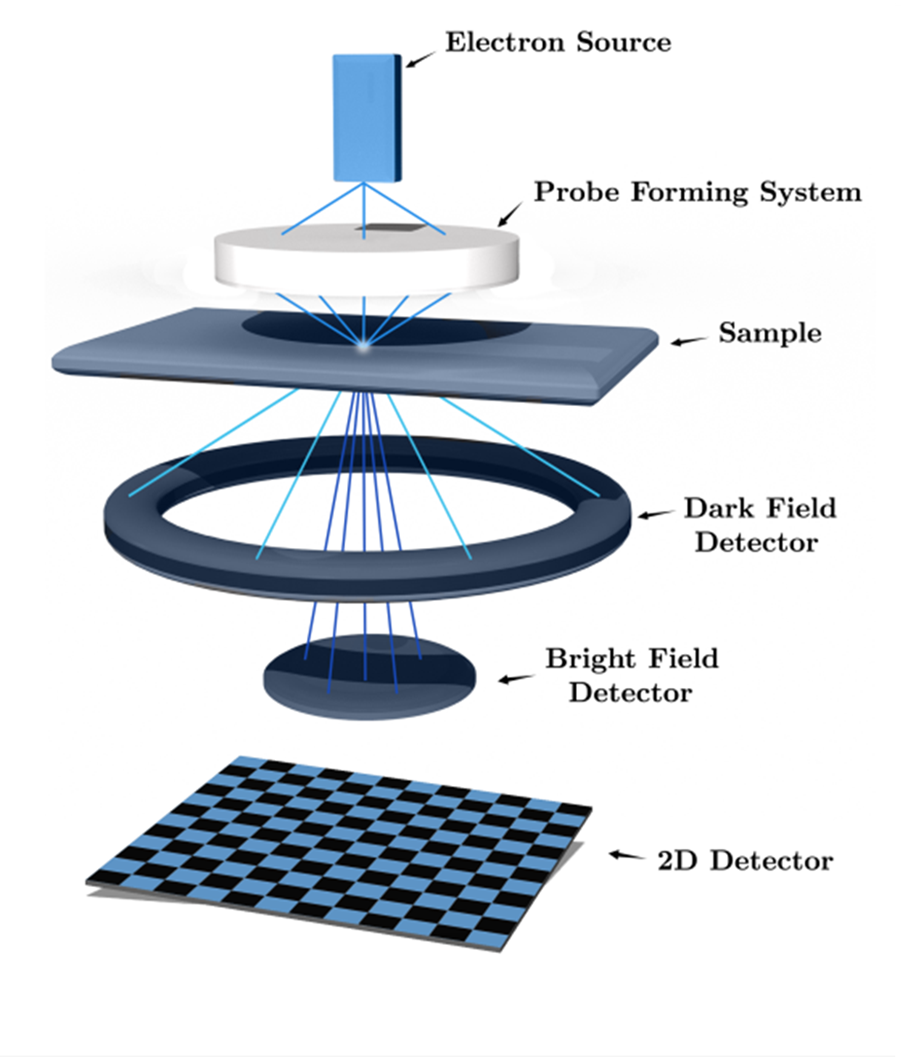}
    \caption{\textbf{Operating principles of 4-D STEM.} An electron probe scans different locations over a field of view and a diffraction pattern is collected in the far-field regime using a 2-D direct electron detector. This creates a 4-D dataset. Dark field and bright field imaging as in standard STEM are accessible using a subset of scattered electrons which can be computationally defined post acquisition.}
    \label{fig:4d-stem-schematic}
\end{figure}

In conventional STEM, as simply illustrated in Fig.~\ref{fig:4d-stem-schematic}, a convergent electron beam is raster scanned across the sample. Scattered or transmitted electrons can be collected by employing an annular or a disc shape detector, for dark field and bright field signals, respectively. These detectors are conventionally based on a scintillator/photomultiplier layout. Modern implementations use a 2-dimensional (2-D) pixelated detector located at a far field diffraction plane to collect all the forward electrons and to record a 2-D diffraction pattern for every position of the probe; hence, a 4-D STEM dataset for all probe positions. This can then be computationally processed to select the signal of interest by simply masking appropriate portions of the full pixel array. See~\cite[Ch.~17]{hawkes2019springer} for a detailed description.

More specifically, complex imaging techniques such as center of mass (COM) \cite{MullerCASPARY_COM} can be implemented by measuring the momentum change of the probe in the detector plane as the result of the interaction with the electrostatic potential of the sample. Ptychography also makes use of the intensities in the diffraction patterns to retrieve the phase information of the object transmission wave function using either non-iterative retrieval algorithms ~\cite{rodenburg1992theory,rodenburg1993experimental} or various iterative numerical approaches~\cite{maiden2009improved,maiden2024wasp,thibault2009probe}.

Some of the constraints of this configuration are the small field of view and the acquisition speed compared to conventional STEM imaging. The frame readout speed of the detector, which is in the order of milliseconds, limits the time spent by the probe at each position (or dwell time), which is in the order of microseconds in conventional STEM. The increased total acquisition time inherently enhances the adverse effects of sample drift and beam damage. Utilising event-based detectors~\cite{Jannis2022_4DSTEM} or subsampling probe positions~\cite{robinson2024real} are examples of approaches to overcome the acquisition time limitation in ptychography.    

The use of a defocused probe at the sample plane enables scanning of a larger Field of View (FoV) with a reasonable acquisition time by increasing the scan step size, subject to a minimum overlap constraint between probe positions. For reference, the 4-D STEM dataset of Rotavirus~\cite{zhou2020lowdose} used in this paper, which was acquired in approximately $16$s, yields a FoV of approximately $400{\rm nm} \times 400{\rm nm}$. Phase information can be retrieved by using iterative methods~\cite{rodenburg2004phase,thibault2009probe,thibault2012maximum,maiden2009improved}, given sufficient overlap between adjacent illuminated areas \cite{bunk2008influence,moshtaghpour2025overlap}. The electron fluence is reduced on account of the increased scan step size and larger probe size. In this work, we centre our attention on defocused ptychography discussed by Zhou \etal~\cite{zhou2020lowdose}, which has been demonstrated as a suitable method for imaging extremely radiation sensitive biological samples.

The extended Ptychographical Iterative Engine (ePIE)~\cite{maiden2009improved} is a phase retrieval algorithm that recovers the complex object function by making estimations of the exit wave function (describing the product of the probe and object wave functions) propagated to the far-field detector plane where its amplitude is replaced by the square root of the measured intensities. The updated exit wave is then back propagated to the object plane where a new estimation of the object function is made. 

In this paper, we introduce an $\ell_0$ regularisation of the ePIE algorithm~\cite{maiden2009improved}, or in short \algname. We note that in strict mathematical terms, $\ell_0$ defined as $\| \bs u \|_0 \coloneqq |\{i: |u_i| \ne 0\}|$, which only counts the number of non-zero elements, is not a norm nor a pseudo-norm. The $\ell_0$-quasi-norm regularisation, enforces the estimation of the object wave function to have a sparse representation -- \ie only a few coefficients are non-zero -- in a transform domain, \eg in a Discrete Cosine Transform (DCT)~\cite{Strang1999discrete}, Discrete Fourier Transform (DFT), or a Discrete Wavelet Transform (DWT)~\cite{mallat1999wavelet}. The underlying mechanism in $\ell_0$ regularisation is that natural images, \eg those of interest here, can be distinguished from noise-like images based on their sparsity level in a transform domain. This principle has been leveraged by a large number of denoising and signal recovery algorithms~\cite{donoho1995denoising,taswell2000and} and in the theory of compressive sensing~\cite{donoho2006compressed,candes2006stable}. In the context of this paper, $\ell_0$ regularisation of the estimated object wave function reduces to only three computationally scalable steps: \textit{(i)} transforming the phase and amplitude of the estimated object wave function to a suitable domain, which is a matter of choice or design, \textit{(ii)} an application of a hard-thresholding operator with parameter $K$, which maps all but $K$ (absolute value) largest coefficients to zero, and \textit{(iii)} transforming the processed coefficients back to the original domain. By artificially down-sampling the probe positions on experimental 4-D STEM data of double layered Rotavirus particles acquired with
85\% overlap, we show \algname is more robust to a reduced overlap ratio compared to conventional ePIE. These  suggest that using this approach, the fluence delivered to the sample could be reduced by scanning larger areas without compromising the quality of the reconstructed object wave function compared. Preliminary results of this work were presented in \cite{moshtaghpour2022towards,moshtaghpour2023exploring}.
%=============== New Section ==============
\section{4-D STEM acquisition model and electron wave propagation}\label{sec:4d-stem}
In this section, a summary of the fundamentals of a 4-D STEM acquisition and associated electron wave propagation are described within a simplified discrete mathematical model.  

Consider a STEM scan with \textit{(i)} a 2-D pixelated detector with $\Hd \times \Wd$ pixels and \textit{(ii)} a scanning probe programmed to visit $\Hp \times \Wp$ locations over the object. In 4-D STEM, a 2-D diffraction pattern is recorded for every 2-D probe location.

Let $\bs o \in \bb C^{N_{\rm o}}$ be the discretised and vectorised version of the complex-valued object wave function over $\Ho \times \Wo$ regularly-spaced points in the real-space with $N_{\rm o} \coloneqq H_{\rm o}W_{\rm o}$. Similarly, let $\bs p \in \bb C^{N_{\rm d}}$ with $\Nd \coloneqq \Hd \Wd$ be the discretised and vectorised version of the complex-valued probe wave function over $\Hd \times \Wd$ regularly-spaced points in the real-space. Therefore, for every probe position indexed by $l \in \dbracket{\Np} \coloneqq \{1,\cdots, \Np\}$, with $\Np \coloneqq \Hp \Wp$ total number of probe locations, the exit wave function of an object where a multiplicative relationship between probe and object holds is
\begin{equation}\label{eq:exit-wave-equation}
    \textrm\quad\bs \psi_l = \bs p \odot (\bs \Pi_l \bs o) \in \bb C^{\Nd}, \quad {\rm for~} l \in \dbracket{\Np},
\end{equation}
where $\odot$ is the Hadamard (or element-wise) product and $\bs \Pi_{l} \in \{0,1\}^{\Nd\times \No}$ is a binary matrix such that $\bs \Pi_l \bs o$ selects only entries of $\bs o$ that are inside the region illuminated by the probe at a location indexed by $l$.

Assuming that data is collected in the far-field or Fraunhofer plane, the wave function at the detector plane, can be modelled as the Fourier transform of the exit wave~\cite[Ch.~17]{hawkes2019springer}, \ie
\begin{equation}\label{eq:detector-wave-equation}
    \bs \phi_l = \bs F \bs \psi_l \in \bb C^{\Nd}, \quad {\rm for~} l \in \dbracket{\Np},
\end{equation}
where $\bs F \in \bb C^{\Nd \times \Nd}$ is the 2-D Discrete Fourier Transform (DFT) matrix. 
Finally, diffraction patterns are acquired by the detector as intensities \ie
\begin{equation}\label{eq:diffraction-pattern-equation}
    \textrm\quad\bs y_l = \left|{\bs \phi}_l \right|^2 \in \bb R^{\Nd}, \quad {\rm for~} l \in \dbracket{\Np},
\end{equation}
where the operator $\left|\bs \cdot \right|^2$ is element-wise. Therefore, combining Eqs.~\eqref{eq:exit-wave-equation}, \eqref{eq:detector-wave-equation}, and \eqref{eq:diffraction-pattern-equation} yields a simplified sensing (or forward) model for 4-D STEM: for every probe position index $l\in \dbracket{\Np}$, 
\begin{equation}\label{eq:4d-stem-equation}
    \textrm\quad\bs y_l = 
    \left|\bs F (\bs p \odot \bs \Pi_l \bs o) \right|^2 + \bs n_l,
\end{equation}
where $\bs n_l \in \bb R^{\Nd}$ is a noise model appropriate for the detector. We refer the reader to~\cite{pix_detectorsI} for practical noise models appropriate for direct electron detectors.

%-------------- New subsection --------------
\section{Ptychography methods: from ePIE to regularised ePIE}\label{sec:ptychography}

Our regularised algorithm is based on the ePIE algorithm. Therefore, in this section, we first provide a suitable mathematical formulation of ePIE followed by an adaptation that yields its $\ell_0$ regularised version. We note that the proposed $\ell_0$ regularisation can be directly applied to ptychography in the near-field or Fresnel regime.

The ePIE algorithm~\cite{maiden2009improved} is an iterative method -- based on the principles of the Gerchberg-Saxton~\cite{gerchberg1971phase} and Fienup~\cite{fienup1982phase} phase retrieval algorithms -- for joint estimation of complex-valued object and probe data. Given a 4-D STEM measurements described by \eqref{eq:4d-stem-equation} and an initial guess of the object and probe data, ePIE performs multiple iterations over the whole dataset. 

As shown in Algorithm ~\ref{alg:epie}, ePIE takes as input the 4-D STEM data $\{\bs y_{l}\}_{l=1}^{\Np}$, initial guesses for object $\bs o^{(0)}$ and probe $\bs p^{(0)}$, a number of iterations $N_{\rm itr}$, and update step size parameters for the object $\alpha_{\rm o}$ and probe $\alpha_{\rm p}$. An initial probe estimate can be generated based on the experimentally measured aberration parameters, and a common approach to  initialise object data is to assume unit amplitude with random phase components. 

\begin{algorithm}[!t]
\caption{ePIE and LoRePIE algorithms for electron ptychography}\label{alg:epie}
\KwInput{$\{\bs y_l\}_{l=1}^{\Np}, \bs o^{(0)},  \bs p^{(0)} \in \bb C^{\Nd}, N_{\rm itr}, \alpha_{\rm o}, \alpha_{\rm p}$}
\KwOutput{$\hat{\bs o} = \bs o^{(N_{\rm itr})}, \hat{\bs p} = \bs p^{(N_{\rm itr})}$}
\For{$t \in \llbracket N_{\rm itr}\rrbracket$}{
$\bs o^{(t)}  \leftarrow \bs o^{(t-1)}, \bs p^{(t)}  \leftarrow \bs p^{(t-1)}$\\
\For{$l \in \llbracket \Np \rrbracket$}{
$\bs {o}^{(t)}_l  \leftarrow  \bs \Pi_l \bs o^{(t)}$\tcp*[f]{Cropped object FoV}\\
$\bs \psi^{\rm e}_l  \leftarrow \bs p^{(t)} \odot  \bs o^{(t)}_l
$\tcp*[f]{Estimated exit wave function}\\
$\bs \phi^{\rm e}_l  \leftarrow \bs F \bs \psi^{\rm e}_l$\tcp*[f]{Estimated wave function at detector plane}\\
$\bs \phi^{\rm u}_l   \leftarrow \sqrt{\bs y_l}\odot e^{i\angle\,\bs \phi^{\rm e}_l}$\tcp*[f]{Updated wave function at detector plane}\\
$\bs \psi^{\rm u}_l  \leftarrow \bs F^{-1} \bs \phi^{\rm u}_l
$\tcp*[f]{Updated exit wave function}\\
$\bs p^{(t)} \leftarrow \bs p^{(t)} + \alpha_{\rm p}\frac{\bs o_l^{(t)*}}{\| \bs o^{(t)}_l\|^2_{\infty}}\odot (\bs \psi^{\rm u}_l - \bs \psi^{\rm e}_l)$\tcp*[f]{Updated probe data}\\
$\bs o^{(t)}_l  \leftarrow \bs o^{(t)}_l + \alpha_{\rm o}\frac{\bs p^{(t)*}}{\| \bs p^{(t)}\|^2_{\infty}}\odot (\bs \psi^{\rm u}_l - \bs \psi^{\rm e}_l)$\tcp*[f]{Updated Object data}\\
$\bs o^{(t)}_l  \leftarrow \cl D(\bs o^{(t)}_l)$\tcp*[f]{(Only in LoRePIE) Regularised object data}\\
$\bs o^{(t)}  \leftarrow \bs \Pi^\top \bs o^{(t)}_l + (\bs I_{\No} - \bs \Pi^\top\bs \Pi) \bs o^{(t)}$\tcp*[f]{Full object data}
}
}
\end{algorithm} 

At every ePIE iteration $t \in \dbracket{N_{\rm itr}}$, and for a randomly selected probe position index $l\in \dbracket{\Np}$, a corresponding illuminated region of the full object is first computed (Line 4). Based on Eq.~\eqref{eq:exit-wave-equation}, an estimated exit wave is next computed (Line 5). According to Eq.~\eqref{eq:detector-wave-equation}, the wave function at the detector plane is computed by an application of the DFT (Line 6).
The next step of ePIE is to provide an updated estimate of this wave function by replacing its amplitude with the square root of the recorded diffraction pattern (Line 7). In Line 7, both $\sqrt{.}$ 
 and $\angle\, \cdot$ are element-wise and $\angle\,\bs \phi^{\rm e}_l$ returns the phase of $\bs \phi^{\rm e}_l$. The update rule in Line 7 enforces a consistency between the amplitude of the wave function at the detector plane and the square root of the recorded intensities. In the next step (Line 8), an updated exit wave $\bs \psi^{\rm u}_l$ is computed by an application of the inverse DFT on the updated wave function in the detector plane.

 The ePIE iteration continues with new estimates of probe $\bs p^{(t)}$ (Line 9) and object $\bs o^{(t)}_l$ (Line 10) such that $\bs p^{(t)} \odot \bs o^{(t)}_l = \bs \psi^{\rm u}_l$. Since simultaneous estimation of $\bs p^{(t)}$ and $\bs o^{(t)}_l$ is not feasible, those estimates are performed in ePIE in an alternating manner, using the previous object data for updating the probe data and using the new probe data for updating the object data. In Lines 9 and 10 of Algorithm.~\ref{alg:epie}, $\|\bs u\|_\infty \coloneqq \max_j |u_j|$ returns the maximum absolute value of the vector $\bs u$. The last step of the ePIE iteration is to put back the new estimate of the object crop in the full FoV of object (Line 12).

 Despite its simplicity and elegance, ePIE is prone to recover low quality images when the probe overlap is low. The authors of \cite{bunk2008influence} advised an overlap of 60\% as a reasonable trade-off between recovery quality and radiation dose. Additionally, ePIE does not take into account an explicit noise model, as the update rule in Line 7 of Algorithm~\ref{alg:epie} implicitly assumes the diffraction patterns are noise-free. Finally, we note that ePIE is also prone to stagnation under certain conditions, including when recovering data under low fluence conditions 

We now introduce $\ell_0$ regularised ePIE that is robust to low probe overlap ratios. The underlying idea of \algname is to regularise the estimation of object crops $\bs o^{(t)}_l$ assuming that both its phase and amplitude have a sparse representation in a transform domain. See Sec.~S1 in Supplementary Material for an illustration of this idea. As outlined in Algorithm~\ref{alg:epie}, LoRePIE involves the same steps as those in ePIE algorithm, with an additional step (Line 11) that regularises the object data using a function $\cl D: \bb C^{\Nd} \mapsto \bb C^{\Nd}$. 

For a vector $\bs u \in \bb C^{N}$, function $\cl D$  takes the form of
\begin{equation}\label{eq:lorepie-D}
\cl D(\bs u) \coloneqq \cl D_{\rm amp}(|\bs u|) \odot \exp(i \cl D_{\rm phs}(\angle \bs u)),
\end{equation}
where $\cl D_{\rm amp}: \bb R^{N} \mapsto \bb R^{N}$ and $\cl D_{\rm phs}: \bb R^{N} \mapsto \bb R^{N}$ are general regularisation functions associated with, respectively, the amplitude and phase of the object data. In this paper, we consider $\ell_0$-denoising function for both amplitude and phase regularisation represented, respectively, as
\begin{align}\label{eq:lorepie-DX}
\cl D_{\rm amp}(\bs u) &\coloneqq \bs A_{\rm amp}^{-1} \cl H_{K_{\rm amp}}(\bs A_{\rm amp} \bs u),\\
\cl D_{\rm phs}(\bs u) &\coloneqq \bs A_{\rm phs}^{-1} \cl H_{K_{\rm phs}}(\bs A_{\rm phs} \bs u),
\end{align}
where $\bs A_{\rm amp} \in \bb C^{N \times N}$ and $\bs A_{\rm phs} \in \bb C^{N \times N}$ are assumed here to be a matrix representation of orthogonal transformations, \eg DCT, DFT, or DWT. Given an integer regularisation parameter $K$, the hard-thresholding function $\cl H_K(\bs v)$  maps all but $K$ largest (in absolute value) elements of its input $\bs v$ to zero. We note that this construction of the function $\cl D$ is not arbitrary. Indeed, from Eq.~\eqref{eq:lorepie-D}, it can be shown that the regularised object update in Line 11 of Algorithm~\ref{alg:epie} is the solution to the following minimisation problem:
\begin{align}
\argmin_{\bs u \in \bb C^{\Nd}} & \||\bs u| - |\bs o^{(t)}_l| \|_2^2  + \|\angle\bs u - \angle \bs o^{(t)}_l \|_2^2 
 \label{eq:lorepie-minimisation} \\
& {\rm subject~to}\quad \| \bs A_{\rm amp} |\bs u| \|_0 \le K_{\rm amp}, \nonumber\\
& \phantom{\rm subject~to}\quad \|\bs A_{\rm phs} \angle \bs u \|_0 \le K_{\rm phs},\nonumber
\end{align}
which ensures that both amplitude and phase of the object data are the closest candidates to those of $\bs o^{(t)}_l$ -- in the $\ell_2$-norm sense, with $\| \bs u \|_2 \coloneqq (\sum_j u_j^2)^{1/2}$ --  such that both amplitude and phase components have fewer than, respectively, $K_{\rm amp}$ and $K_{\rm phs}$, non-zero elements in their corresponding transform domains.

In comparison to ePIE, \algname requires four design parameters: two regularisation parameters and two sparsity bases. The proposed regularisation modification to the ePIE algorithm does not cause divergence of the algorithm. However, the choice of regularisation parameters, $K_{\rm phs}$ and $K_{\rm amp}$, influences the quality of the recovered object. For an orthogonal sparsity basis, two extreme cases can be considered: (1) when $K_{\rm phs} = K_{\rm amp} = N_{\rm d}$, no regularisation is applied, and LoRePIE reduces to ePIE; (2) when $K_{\rm phs} = K_{\rm amp} = 1$, LoRePIE estimates an object with constant amplitude and phase. For sufficiently large regularisation parameters, determined by the signal-to-noise ratio, LoRePIE outperforms ePIE.

The applicability of the proposed regularisation depends on the nature of the object and the choice of sparsity basis. Unlike total variation regularisation, our $\ell_0$ denoiser does not promote piecewise smoothness, making it less suitable for objects with highly structured features. Additionally, the choice of the sparsity domain significantly affects performance; for instance, enforcing sparsity in the DCT or DFT domains is beneficial for natural images with periodic features~\cite{mallat1999wavelet}, but may not be optimal for highly oscillatory or amorphous structures. An additional consideration is parameter selection: excessively small values of $K_{\rm phs}$ and $K_{\rm amp}$ may suppress information, while large values reduce the regularisation effect. In principle, low quality, measured data requires stronger regularisation or, equivalently, smaller values for $K_{\rm amp}$ and $K_{\rm phs}$. Investigating alternative sparsity domains, such as DWT, and adaptive parameter selection strategies is ongoing.
 
We note that related versions of an iterative hard thresholding algorithm~\cite{blumensath2008iterative,blumensath2009iterative} have been reported and together with denoising using hard thresholding \cite{donoho1995denoising}. More specifically, the object regularisation step of \algname in Line 11 of Algorithm~\ref{alg:epie} can also be viewed as a denoising of the estimated object data. 

The choice of $\ell_0$ regularisation in our approach is motivated by its simplicity and effectiveness as a non-linear filtering technique. Unlike low-pass filters, which indiscriminately suppress high-frequency Fourier components, $\ell_0$ regularisation selects coefficients in a transform domain, Fourier or otherwise, based on their intensity, thereby preserving fine details while reducing noise. Applying $\ell_0$ regularisation separately to the amplitude and phase components of an object estimate offers flexibility in handling these independently. For instance, in the case of a phase-only object, a natural choice is to set $K_{\rm amp} = 1$, which enforces a constant amplitude regardless of the regularisation of the phase component. This independent treatment of phase and amplitude also facilitates extensions to alternative regularisation approaches, such as enforcing sparsity in the gradient domain or replacing the $\ell_0$ quasi-norm with $\ell_1$ norm. These extensions are a focus of ongoing investigations.

Similar regularisations have been incorporated in ptychographic algorithms. For example, recently a half-quadratic splitting framework has been used to apply regularisations in the form of plug-and-play priors~\cite{denker2024plug}, and a quadratic regularisation for optical ptychography has been explored in~\cite{zach2025perturbative} and reference \cite{schloz2020overcoming} proposed a conjugate gradient descent algorithm regularised by an $\ell_1$ norm for electron ptychography. In the field of Fourier ptychography, where the object is illuminated using plane wave tilted illumination, as opposed to a convergent illumination in electron ptychography, regularisation techniques have been extensively applied in the object estimation process. Notable examples include total variation and Tikhonov~\cite{lee2024anisotropic}, promoting $\ell_0$ sparsity in Block-Matching and 3D (BM3D)~\cite{dabov2007image} frames~\cite{zhang2017fourier}, and image denoisers~\cite{sun2019regularized}.

%=============== New Section ==============
\section{Numerical results}\label{sec:numerical-results}
\begin{figure}[tb]
    \centering
    \scalebox{1}{\includegraphics[width=0.78\columnwidth]{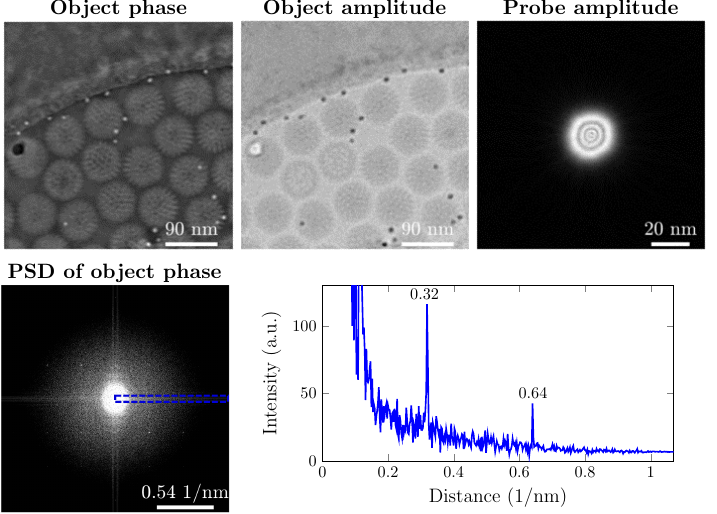}}
    \caption{\textbf{Reference data used for evaluation.} These data are generated by 100 iterations of the ePIE algorithm followed by another 100 iterations of the DM algorithm. The periodic features in the PSD of the object data are due to the grid artefacts, since inverting the frequency of the first periodic peak in line profile curve gives the scan step size: $1/0.32~{\rm nm}^{-1} = 3.125~{\rm nm}$.}
    \label{fig:reference_data}
\end{figure}
In this section, we describe simulations for the application of \algname on both fully sampled experimental data and artificially down-sampled data that are equivalent to datasets with various overlap ratios. 

\sqp\noindent \textbf{Fully acquired experimental data.} We use a 4-D STEM dataset of Rotavirus double-layered particles published in~\cite{zhou2020lowdose}. We refer the reader to \cite{zhou2020lowdose} for the experimental data acquisition details. This dataset was acquired with $(\Hp, \Wp) = (127, 127)$ or $\Np = 127^2$ probe positions, $(\Hd, \Wd) = (256, 256)$ or $\Nd = 256^2$ detector pixels, a scan step size $\Delta_{\rm p} = 3.125$ nm, a convergence semi-angle of $1.034$ mrad, a defocus of $-1.3~\mu$m, and a fluence of $22.8~{\rm e^{-}}/\text{\r{A}}^{2}$. This setting amounts to an overlap percentage of $85\%$. Following reports in the X-ray ptychography literature~\cite{edo2013sampling,da2015elementary,batey2014reciprocal}, we also compute a ptychographic oversampling ratio as defined in \cite{batey2014reciprocal} as $\lambda /2\Delta_{\rm k} \Delta_{\rm p}$, where $\lambda$ is wavelength and $\Delta_{\rm k}$ is angular size of the detector pixel. The ptychographic oversampling ratio corresponding to the fully acquired dataset used here, where $\lambda = 1.9687$~pm and $\Delta_{\rm k} = 0.0165$~mrad, equals $19.05$.

\begin{figure*}[bth]
    \centering
    \scalebox{1}{\includegraphics[width=\linewidth]{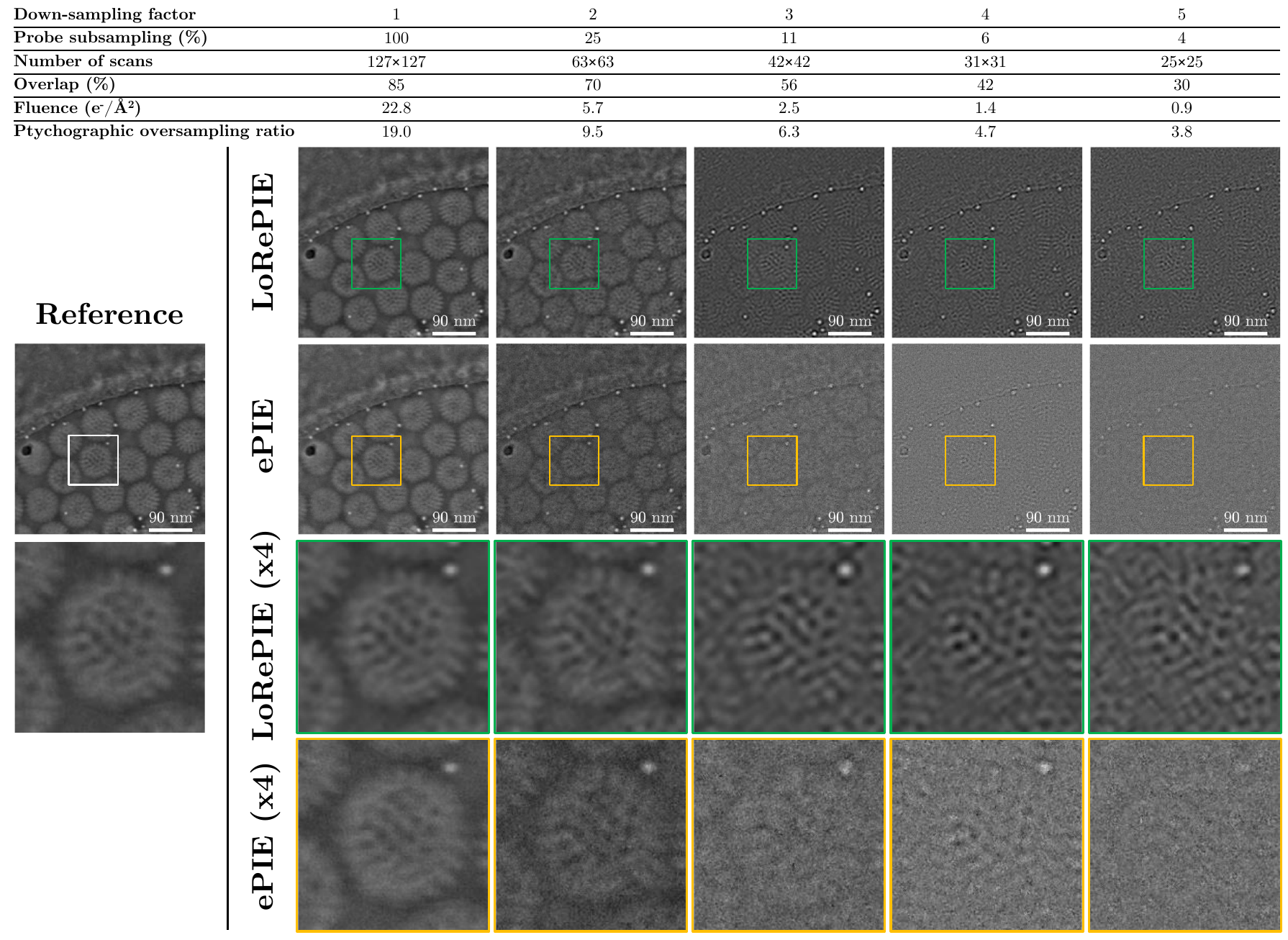}}
    \caption{Reference and reconstructed object phase from LoRePIE and ePIE for various probe position down-sampling factors. LoRePIE consistently achieves higher-quality reconstructions for all down sampling factors, demonstrating improved robustness to reduced probe overlap.}
    \label{fig:main_examples}
\end{figure*}

\sqp\noindent \textbf{Simulated down-sampled data.} A set of datasets with varying overlap ratios were simulated by artificially down-sampling (or uniformly subsampling) the probe positions. As shown in Fig.~\ref{fig:main_examples}, each down-sampling ratio in $\{1,2,3,4,5\}$ corresponds to a value for the subsampling ratio, the number of scanned probe positions, the probe overlap, fluence, and the ptychographic over-sampling ratio. Despite their general feasibility, performing other subsampling strategies of probe positions according to, \eg a uniformly at random or linehop scheme, is not discussed here, since these strategies amount to a non-uniform distribution of both electron dose and overlap. 

\sqp\noindent \textbf{Reference data.} To evaluate the performance of LoRePIE against ePIE and to avoid bias during comparison of the reconstructions generated, the reference object and probe data were obtained from 100 iterations of the ePIE algorithm followed by 100 iterations of the Difference Map (DM)~\cite{thibault2009probe} algorithm utilising the well-established Ptypy~\cite{enders2016computational} package. Subsequent application of ptychographical algorithms is a typical routine performed, \eg in~\cite{tadesse2019wavelength,daurer2021ptychographic,wakonig2020ptychoshelves}, for avoiding stagnation of a phase retrieval process stuck at a local minimum. For the ePIE algorithm, the object and probe update step size parameters were set to $\alpha_{\rm o} = \alpha_{\rm p} = 0.01$. For the DM algorithm, both the object and probe inertia parameters are set to 0.001 and the mix parameter between DM and alternating projection is set to 0.75. 

Fig.~\ref{fig:reference_data} shows the amplitude and phase components of the reference object and the amplitude of probe data. The Power Spectrum Density (PSD) of the object phase reveals periodic frequency features. By examining the line profile of the object phase in the PSD map, we note that these periodic features are due to the grid artefacts \cite{HUE2011ptychoSTEM}, since inverting the frequency of the first periodic peak in Fig.~\ref{fig:reference_data}-bottom gives the scan step size: $1/0.32~{\rm nm}^{-1} = 3.125~{\rm nm}$. 

\begin{figure}[!ht]
    \centering
    \scalebox{0.95}{\includegraphics[width=0.5\textwidth]{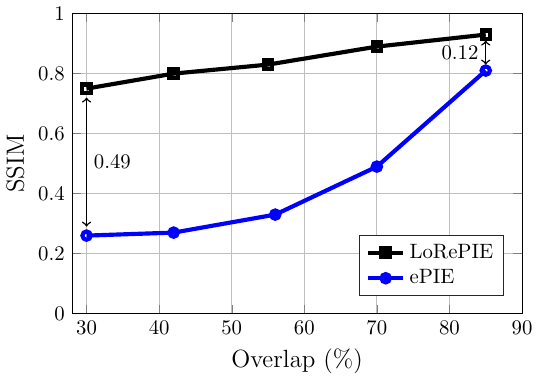}}
    \caption{Performance comparison between LoRePIE and ePIE for a range of overlap percentages, determined by down-sampling the probe positions. Reference and reconstructed object phase data are shown in Fig.~\ref{fig:main_examples}.}
    \label{fig:ssim_overlap_curve_lorepie_epie}
\end{figure}

\sqp\noindent \textbf{Ablation studies.} The hyperparameters were fine tuned as described in the Supplementary Material, Secs.~S2, S3, and S4. Various values of the regularisation parameters are tested in Sec.~S2,  which shows that setting $K_{\rm amp} = K_{\rm phs} = 0.05 \Nd$ yields the highest NRMSE values for all down-sampling factors. Based on our observations in Sec.~S3, using 50 iterations of the LoRePIE and ePIE algorithms offers a reasonable trade-off between reconstruction quality and computational time. For the range of probe and object update step size parameters given in Sec.~S4, we investigated efficient values of these parameters for the ePIE and LoRePIE algorithms for each down-sampling factor. For all subsequent reconstructions, we use the update step size values that yields the highest Structural Similarity Index Measure (SSIM) ~\cite{wang2004image} values in Fig.~S12.

\sqp\noindent \textbf{Results.}
Fig.~\ref{fig:main_examples} shows the reference and reconstructed object phase data from the LoRePIE and ePIE algorithms. SSIM values between the reference and reconstructed object phase data are calculated in Fig.~\ref{fig:ssim_overlap_curve_lorepie_epie} for a range of down-sampling factors, which correspond to varying overlap percentages. The LoRePIE algorithm demonstrates better robustness to decreasing overlap and consistently outperforms the ePIE algorithm. In contrast, ePIE produces significantly lower SSIM values when the probe positions are down-sampled (Fig.~\ref{fig:main_examples}). Visual inspection indicates that LoRePIE gives a slightly higher quality -- in the sense of reduced noise elements -- in the reconstructed object phase compared to the reference data, suggesting a need to further investigate the LoRePIE's performance using synthetic 4-D STEM datasets. The ePIE algorithm fails to achieve sufficient quality in object phase reconstruction when probe positions are down-sampled by a factor greater than two. In contrast, LoRePIE enables reconstruction with good visual quality at a down-sampling factor of three. Even at down-sampling factors of four or five, corresponding to very low overlap percentages, LoRePIE produces reconstructed object phase images of sufficient quality to perform some useful qualitative assessments, such as counting the number of particles or measuring their size.
%=============== New Section ==============
\section{Conclusion}\label{sec:conclusion}
In this work, we have introduced an $\ell_0$-regularised version of the ePIE algorithm (LoRePIE) which demonstrated enhanced robustness to reduced overlap ratios between illuminated areas in 4-D STEM experiments. LoRePIE applies $\ell_0$-denoising to regularise the estimated object data, enforcing that both the amplitude and phase of the object are sparsely represented in a transform domain. This was achieved by application of a hard-thresholding operator on the coefficients in that transform domain. We have compared LoRePIE to ePIE  using a 4-D STEM dataset of Rotavirus particles taken from~\cite{zhou2020lowdose}. Various versions of the dataset were created to simulate different overlap percentages by down-sampling the probe positions. Our results indicate that LoRePIE consistently outperforms ePIE in reconstructing object phases, on both fully acquired data and down-sampled data with low overlap percentages.

In this work, the Discrete Cosine Transform (DCT) domain was utilised as the basis for thresholding the coefficients of the object phase and amplitude.  Future studies will explore the performance of LoRePIE with alternative sparsifying bases, such as wavelet transforms.
%=============== New Section ==============
\section{Acknowledgement}\label{sec:acknowledgement}
The authors thank Jack D.~Wells for designing Fig.~\ref{fig:4d-stem-schematic} Peng Wang, Zhiyuan Ding, and Chen Huang for providing the 4-D STEM dataset used in this paper and for assistance in reproducing the results of \cite{zhou2020lowdose}. Benedikt J.~Daurer and other developers of Ptypy package~\cite{enders2016computational} are acknowledged for assisting the authors with the simulations. The authors also thank Andrew Maiden, Laurent Jacques, and David A.~Muller for  valuable comments during the manuscript preparation. 
\subsection{Funding}
Rosalind Franklin Institute; UK Research and Innovation; Engineering and Physical Sciences Research Council; University of Liverpool; University of Oxford.
\subsection{Disclosures}
The authors declare no conflicts of interest.

\subsection{Data availability} The data that support the findings of this study are available from the corresponding author upon reasonable request.

\subsection{Supplemental document}
See the supplementary material for ablation studies conducted to fine-tune the hyperparameters of the ePIE and 
LoRePIE algorithms.
%=============== New Section ==============
\bibliographystyle{IEEEtran}
\bibliography{Main}

% Generated by IEEEtran.bst, version: 1.14 (2015/08/26)
\begin{thebibliography}{10}
\providecommand{\url}[1]{#1}
\csname url@samestyle\endcsname
\providecommand{\newblock}{\relax}
\providecommand{\bibinfo}[2]{#2}
\providecommand{\BIBentrySTDinterwordspacing}{\spaceskip=0pt\relax}
\providecommand{\BIBentryALTinterwordstretchfactor}{4}
\providecommand{\BIBentryALTinterwordspacing}{\spaceskip=\fontdimen2\font plus
\BIBentryALTinterwordstretchfactor\fontdimen3\font minus \fontdimen4\font\relax}
\providecommand{\BIBforeignlanguage}[2]{{%
\expandafter\ifx\csname l@#1\endcsname\relax
\typeout{** WARNING: IEEEtran.bst: No hyphenation pattern has been}%
\typeout{** loaded for the language `#1'. Using the pattern for}%
\typeout{** the default language instead.}%
\else
\language=\csname l@#1\endcsname
\fi
#2}}
\providecommand{\BIBdecl}{\relax}
\BIBdecl

\bibitem{krivanek1999towards}
O.~Krivanek, N.~Dellby, and A.~Lupini, ``Towards sub-{\aa} electron beams,'' \emph{Ultramicroscopy}, vol.~78, no. 1-4, pp. 1--11, 1999.

\bibitem{batson2002sub}
P.~E. Batson, N.~Dellby, and O.~L. Krivanek, ``Sub-{\aa}ngstrom resolution using aberration corrected electron optics,'' \emph{Nature}, vol. 418, no. 6898, pp. 617--620, 2002.

\bibitem{HAADFHowie1979}
A.~Howie, ``Image contrast and localized signal selection techniques,'' \emph{Journal of Microscopy}, vol. 117, pp. 11--23, 2 1979.

\bibitem{HAADFPennycook1989}
S.~J. Pennycook, ``Z-contrast {STEM} for materials science,'' \emph{Ultramicroscopy}, vol.~30, pp. 58--69, 12 1989.

\bibitem{PENNYCOOK201722}
------, ``The impact of {STEM} aberration correction on materials science,'' \emph{Ultramicroscopy}, vol. 180, pp. 22--33, 2017.

\bibitem{chen2021electron}
Z.~Chen, Y.~Jiang, Y.-T. Shao, M.~E. Holtz, M.~Odstr{\v{c}}il, M.~Guizar-Sicairos, I.~Hanke, S.~Ganschow, D.~G. Schlom, and D.~A. Muller, ``Electron ptychography achieves atomic-resolution limits set by lattice vibrations,'' \emph{Science}, vol. 372, no. 6544, pp. 826--831, 2021.

\bibitem{Bio_STEM2012}
A.~A. Sousa and R.~D. Leapman, ``Development and application of stem for the biological sciences,'' \emph{Ultramicroscopy}, vol. 123, pp. 38--49, 2012.

\bibitem{zhou2020lowdose}
L.~Zhou, J.~Song, J.~S. Kim, X.~Pei, C.~Huang, M.~Boyce, L.~Mendonça, D.~Clare, A.~Siebert, E.~Allen, Christopher S.~Liberti, D.~Stuart, X.~Pan, P.~D. Nellist, P.~Zhang, A.~I. Kirkland, and P.~Wang, ``Low-dose phase retrieval of biological specimens using cryo-electron ptychography,'' \emph{Nature Communications}, vol.~11, p. 2773, 2020.

\bibitem{van2009imaging}
K.~Van~Benthem and S.~J. Pennycook, ``Imaging and spectroscopy of defects in semiconductors using aberration-corrected stem,'' \emph{Applied Physics A}, vol.~96, pp. 161--169, 2009.

\bibitem{chejarla2023measuring}
V.~S. Chejarla, S.~Ahmed, J.~Belz, J.~Scheunert, A.~Beyer, and K.~Volz, ``Measuring spatially-resolved potential drops at semiconductor hetero-interfaces using 4d-stem,'' \emph{Small Methods}, vol.~7, no.~9, p. 2300453, 2023.

\bibitem{EGERTON2004399}
R.~Egerton, P.~Li, and M.~Malac, ``Radiation damage in the tem and sem,'' \emph{Micron}, vol.~35, no.~6, pp. 399--409, 2004.

\bibitem{velazcoreducing2022}
A.~Velazco, A.~Béché, D.~Jannis, and J.~Verbeeck, ``Reducing electron beam damage through alternative {STEM} scanning strategies, part i: Experimental findings,'' \emph{Ultramicroscopy}, vol. 232, p. 113398, 2022.

\bibitem{nicholls2022compressive}
D.~Nicholls, A.~Robinson, J.~Wells, A.~Moshtaghpour, M.~Bahri, A.~Kirkland, and N.~Browning, ``Compressive scanning transmission electron microscopy,'' in \emph{proceedings of 2022 IEEE International Conference on Acoustics, Speech and Signal Processing (ICASSP)}.\hskip 1em plus 0.5em minus 0.4em\relax IEEE, 2022, pp. 1586--1590.

\bibitem{pix_detectorsI}
\BIBentryALTinterwordspacing
B.~D.~A. Levin, ``Direct detectors and their applications in electron microscopy for materials science,'' \emph{Journal of Physics: Materials}, vol.~4, no.~4, p. 042005, 2021. [Online]. Available: \url{https://dx.doi.org/10.1088/2515-7639/ac0ff9}
\BIBentrySTDinterwordspacing

\bibitem{pix_detectorsII}
\BIBentryALTinterwordspacing
I.~MacLaren, T.~A. Macgregor, C.~S. Allen, and A.~I. Kirkland, ``{Detectors—The ongoing revolution in scanning transmission electron microscopy and why this important to material characterization},'' \emph{APL Materials}, vol.~8, no.~11, p. 110901, 2020. [Online]. Available: \url{https://doi.org/10.1063/5.0026992}
\BIBentrySTDinterwordspacing

\bibitem{Ophus_2019}
C.~Ophus, ``Four-dimensional scanning transmission electron microscopy (4d-stem): From scanning nanodiffraction to ptychography and beyond,'' \emph{Microscopy and Microanalysis}, vol.~25, no.~3, p. 563–582, 2019.

\bibitem{song2019atomic}
J.~Song, C.~S. Allen, S.~Gao, C.~Huang, H.~Sawada, X.~Pan, J.~Warner, P.~Wang, and A.~I. Kirkland, ``Atomic resolution defocused electron ptychography at low dose with a fast, direct electron detector,'' \emph{Scientific reports}, vol.~9, no.~1, p. 3919, 2019.

\bibitem{hawkes2019springer}
P.~W. Hawkes and J.~C. Spence, \emph{Springer handbook of microscopy}.\hskip 1em plus 0.5em minus 0.4em\relax Springer Nature, 2019.

\bibitem{MullerCASPARY_COM}
\BIBentryALTinterwordspacing
K.~Müller-Caspary, F.~F. Krause, T.~Grieb, S.~Löffler, M.~Schowalter, A.~Béché, V.~Galioit, D.~Marquardt, J.~Zweck, P.~Schattschneider, J.~Verbeeck, and A.~Rosenauer, ``Measurement of atomic electric fields and charge densities from average momentum transfers using scanning transmission electron microscopy,'' \emph{Ultramicroscopy}, vol. 178, pp. 62--80, 2017. [Online]. Available: \url{https://www.sciencedirect.com/science/article/pii/S0304399116300596}
\BIBentrySTDinterwordspacing

\bibitem{rodenburg1992theory}
J.~M. Rodenburg and R.~H.~T. Bates, ``The theory of super-resolution electron microscopy via wigner-distribution deconvolution,'' \emph{Philosophical Transactions of the Royal Society of London. Series A: Physical and Engineering Sciences}, vol. 339, pp. 521--553, 1992.

\bibitem{rodenburg1993experimental}
J.~M. Rodenburg, B.~C. McCallum, and P.~D. Nellist, ``Experimental tests on double-resolution coherent imaging via stem,'' \emph{Ultramicroscopy}, vol.~48, pp. 304--314, 1993.

\bibitem{maiden2009improved}
A.~M. Maiden and J.~M. Rodenburg, ``An improved ptychographical phase retrieval algorithm for diffractive imaging,'' \emph{Ultramicroscopy}, vol. 109, no.~10, pp. 1256--1262, 2009.

\bibitem{maiden2024wasp}
A.~M. Maiden, W.~Mei, and P.~Li, ``{WASP}: weighted average of sequential projections for ptychographic phase retrieval,'' \emph{Optics Express}, vol.~32, no.~12, pp. 21\,327--21\,344, 2024.

\bibitem{thibault2009probe}
P.~Thibault, M.~Dierolf, O.~Bunk, A.~Menzel, and F.~Pfeiffer, ``Probe retrieval in ptychographic coherent diffractive imaging,'' \emph{Ultramicroscopy}, vol. 109, no.~4, pp. 338--343, 2009.

\bibitem{Jannis2022_4DSTEM}
D.~Jannis, C.~Hofer, C.~Gao, X.~Xie, A.~Béché, T.~Pennycook, and J.~Verbeeck, ``Event driven 4d stem acquisition with a timepix3 detector: Microsecond dwell time and faster scans for high precision and low dose applications,'' \emph{Ultramicroscopy}, vol. 233, p. 113423, 2022.

\bibitem{robinson2024real}
A.~Robinson, J.~Wells, A.~Moshtaghpour, D.~Nicholls, C.~Huang, A.~Velazco-Torrejon, G.~Nicotra, A.~Kirkland, and N.~Browning, ``Real-time four-dimensional scanning transmission electron microscopy through sparse sampling,'' \emph{Chinese Physics B}, vol.~33, no.~11, p. 116804, 2024.

\bibitem{rodenburg2004phase}
J.~M. Rodenburg and H.~M. Faulkner, ``A phase retrieval algorithm for shifting illumination,'' \emph{Applied physics letters}, vol.~85, no.~20, pp. 4795--4797, 2004.

\bibitem{thibault2012maximum}
P.~Thibault and M.~Guizar-Sicairos, ``Maximum-likelihood refinement for coherent diffractive imaging,'' \emph{New Journal of Physics}, vol.~14, no.~6, p. 063004, 2012.

\bibitem{bunk2008influence}
O.~Bunk, M.~Dierolf, S.~Kynde, I.~Johnson, O.~Marti, and F.~Pfeiffer, ``Influence of the overlap parameter on the convergence of the ptychographical iterative engine,'' \emph{Ultramicroscopy}, vol. 108, no.~5, pp. 481--487, 2008.

\bibitem{moshtaghpour2025overlap}
A.~Moshtaghpour and A.~I. Kirkland, ``On overlap ratio in defocused electron ptychography,'' \emph{arXiv preprint arXiv:2502.00762}, 2025.

\bibitem{Strang1999discrete}
G.~Strang, ``The discrete cosine transform,'' \emph{SIAM Review}, vol.~41, no.~1, pp. 135--147, 1999.

\bibitem{mallat1999wavelet}
M.~Stephane, ``A wavelet tour of signal processing,'' 1999.

\bibitem{donoho1995denoising}
D.~L. Donoho, ``De-noising by soft-thresholding,'' \emph{IEEE transactions on information theory}, vol.~41, no.~3, pp. 613--627, 1995.

\bibitem{taswell2000and}
C.~Taswell, ``The what, how, and why of wavelet shrinkage denoising,'' \emph{Computing in science \& engineering}, vol.~2, no.~3, pp. 12--19, 2000.

\bibitem{donoho2006compressed}
D.~L. Donoho, ``Compressed sensing,'' \emph{IEEE Transactions on information theory}, vol.~52, no.~4, pp. 1289--1306, 2006.

\bibitem{candes2006stable}
E.~J. Candes, J.~K. Romberg, and T.~Tao, ``Stable signal recovery from incomplete and inaccurate measurements,'' \emph{Communications on Pure and Applied Mathematics: A Journal Issued by the Courant Institute of Mathematical Sciences}, vol.~59, no.~8, pp. 1207--1223, 2006.

\bibitem{moshtaghpour2022towards}
A.~Moshtaghpour, A.~Velazco-Torrejon, A.~Robinson, E.~Liberti, J.~S. Kim, N.~D. Browning, and A.~I. Kirkland, ``Towards low-dose and fast 4-{D} scanning transmission electron microscopy: New sampling and reconstruction approaches,'' \emph{Microscopy and Microanalysis}, vol.~28, no.~S1, pp. 372--373, 2022.

\bibitem{moshtaghpour2023exploring}
A.~Moshtaghpour, A.~Velazco-Torrejon, A.~W. Robinson, A.~I. Kirkland, and N.~D. Browning, ``Exploring low-dose and fast electron ptychography using l0 regularisation of extended ptychographical iterative engine,'' \emph{Microscopy and Microanalysis}, vol.~29, no. Supplement\_1, pp. 344--345, 2023.

\bibitem{gerchberg1971phase}
R.~W. Gerchberg, ``Phase determination from image and diffraction plane pictures in the electron microscope,'' \emph{Optik}, vol.~34, pp. 275--284, 1971.

\bibitem{fienup1982phase}
J.~R. Fienup, ``Phase retrieval algorithms: a comparison,'' \emph{Applied optics}, vol.~21, no.~15, pp. 2758--2769, 1982.

\bibitem{blumensath2008iterative}
T.~Blumensath and M.~E. Davies, ``Iterative thresholding for sparse approximations,'' \emph{Journal of Fourier analysis and Applications}, vol.~14, pp. 629--654, 2008.

\bibitem{blumensath2009iterative}
------, ``Iterative hard thresholding for compressed sensing,'' \emph{Applied and computational harmonic analysis}, vol.~27, no.~3, pp. 265--274, 2009.

\bibitem{denker2024plug}
A.~Denker, J.~Hertrich, Z.~Kereta, S.~Cipiccia, E.~Erin, and S.~Arridge, ``Plug-and-play half-quadratic splitting for ptychography,'' \emph{arXiv preprint arXiv:2412.02548}, 2024.

\bibitem{zach2025perturbative}
M.~Zach, K.-C. Shen, R.~Cao, M.~Unser, L.~Waller, and J.~Dong, ``Perturbative fourier ptychographic microscopy for fast quantitative phase imaging,'' \emph{arXiv preprint arXiv:2501.07308}, 2025.

\bibitem{schloz2020overcoming}
M.~Schloz, T.~C. Pekin, Z.~Chen, W.~Van~den Broek, D.~A. Muller, and C.~T. Koch, ``Overcoming information reduced data and experimentally uncertain parameters in ptychography with regularized optimization,'' \emph{Optics Express}, vol.~28, no.~19, pp. 28\,306--28\,323, 2020.

\bibitem{lee2024anisotropic}
K.~C. Lee, H.~Chae, S.~Xu, K.~Lee, R.~Horstmeyer, S.~A. Lee, and B.-W. Hong, ``Anisotropic regularization for sparsely sampled and noise-robust fourier ptychography,'' \emph{Optics Express}, vol.~32, no.~14, pp. 25\,343--25\,361, 2024.

\bibitem{dabov2007image}
K.~Dabov, A.~Foi, V.~Katkovnik, and K.~Egiazarian, ``Image denoising by sparse 3-d transform-domain collaborative filtering,'' \emph{IEEE Transactions on image processing}, vol.~16, no.~8, pp. 2080--2095, 2007.

\bibitem{zhang2017fourier}
Y.~Zhang, P.~Song, J.~Zhang, and Q.~Dai, ``Fourier ptychographic microscopy with sparse representation,'' \emph{Scientific reports}, vol.~7, no.~1, p. 8664, 2017.

\bibitem{sun2019regularized}
Y.~Sun, S.~Xu, Y.~Li, L.~Tian, B.~Wohlberg, and U.~S. Kamilov, ``Regularized fourier ptychography using an online plug-and-play algorithm,'' in \emph{ICASSP 2019-2019 IEEE International Conference on Acoustics, Speech and Signal Processing (ICASSP)}.\hskip 1em plus 0.5em minus 0.4em\relax IEEE, 2019, pp. 7665--7669.

\bibitem{edo2013sampling}
T.~Edo, D.~Batey, A.~Maiden, C.~Rau, U.~Wagner, Z.~Pe{\v{s}}i{\'c}, T.~Waigh, and J.~Rodenburg, ``Sampling in x-ray ptychography,'' \emph{Physical Review A}, vol.~87, no.~5, p. 053850, 2013.

\bibitem{da2015elementary}
J.~C. da~Silva and A.~Menzel, ``Elementary signals in ptychography,'' \emph{Optics express}, vol.~23, no.~26, pp. 33\,812--33\,821, 2015.

\bibitem{batey2014reciprocal}
D.~Batey, T.~Edo, C.~Rau, U.~Wagner, Z.~Pe{\v{s}}i{\'c}, T.~Waigh, and J.~Rodenburg, ``Reciprocal-space up-sampling from real-space oversampling in x-ray ptychography,'' \emph{Physical Review A}, vol.~89, no.~4, p. 043812, 2014.

\bibitem{enders2016computational}
B.~Enders and P.~Thibault, ``A computational framework for ptychographic reconstructions,'' \emph{Proceedings of the Royal Society A: Mathematical, Physical and Engineering Sciences}, vol. 472, no. 2196, p. 20160640, 2016.

\bibitem{tadesse2019wavelength}
G.~K. Tadesse, W.~Eschen, R.~Klas, M.~Tschernajew, F.~Tuitje, M.~Steinert, M.~Zilk, V.~Schuster, M.~Z{\"u}rch, T.~Pertsch \emph{et~al.}, ``Wavelength-scale ptychographic coherent diffractive imaging using a high-order harmonic source,'' \emph{Scientific reports}, vol.~9, no.~1, p. 1735, 2019.

\bibitem{daurer2021ptychographic}
B.~J. Daurer, S.~Sala, M.~F. Hantke, H.~K. Reddy, J.~Bielecki, Z.~Shen, C.~Nettelblad, M.~Svenda, T.~Ekeberg, G.~A. Carini \emph{et~al.}, ``Ptychographic wavefront characterization for single-particle imaging at x-ray lasers,'' \emph{Optica}, vol.~8, no.~4, pp. 551--562, 2021.

\bibitem{wakonig2020ptychoshelves}
K.~Wakonig, H.-C. Stadler, M.~Odstr{\v{c}}il, E.~H. Tsai, A.~Diaz, M.~Holler, I.~Usov, J.~Raabe, A.~Menzel, and M.~Guizar-Sicairos, ``Ptychoshelves, a versatile high-level framework for high-performance analysis of ptychographic data,'' \emph{Journal of applied crystallography}, vol.~53, no.~2, pp. 574--586, 2020.

\bibitem{HUE2011ptychoSTEM}
F.~Hüe, J.~Rodenburg, A.~Maiden, and P.~Midgley, ``Extended ptychography in the transmission electron microscope: Possibilities and limitations,'' \emph{Ultramicroscopy}, vol. 111, no.~8, pp. 1117--1123, 2011.

\bibitem{wang2004image}
Z.~Wang, A.~C. Bovik, H.~R. Sheikh, and E.~P. Simoncelli, ``Image quality assessment: from error visibility to structural similarity,'' \emph{IEEE transactions on image processing}, vol.~13, no.~4, pp. 600--612, 2004.

\bibitem{cohen2009compressed}
A.~Cohen, W.~Dahmen, and R.~DeVore, ``Compressed sensing and best k-term approximation,'' \emph{Journal of the American mathematical society}, vol.~22, no.~1, pp. 211--231, 2009.

\end{thebibliography}
\clearpage

%%%%%% End of Title %%%%%%
% \tableofcontents
\section{Supplementary Information}
%%%%%% Start of Section 1 %%%%%%
%%%%%% Start of Section 2 %%%%%%
\subsection{Low-complexity of the Rotavirus phase image}
\label{sec:low_complexity_reference}
The intuition that motivated this work is that natural signals, such as images of Rotavirus, are low-complex in the sense of their sparsity in a transform domain. In the following, we describe our study of the low-complexity of the phase component of the reference Rotavirus image in the Discrete Cosine Transform (DCT) domain. 

Given a phase image $\bs x$ with $N_{\rm d} = H_{\rm d} W_{\rm d}$ pixels and given a sparsity threshold $K_{\rm phs} \in \dbracket{N_{\rm d}}$ we compute the best $K_{\rm phs}$-best term approximation \cite{cohen2009compressed} of $\bs x$ following Eq.~\eqref{eq:lorepie-DX}. The phase image $\bs x$ is first transformed to a 2-D DCT domain. Resulting coefficients are then hard-thresholded such that all but $K_{\rm phs}$ coefficients are mapped to zero followed by an application of the inverse 2-D DCT. 

\begin{figure}[!bth]
    \centering
    \scalebox{1}{\includegraphics[width=\columnwidth]{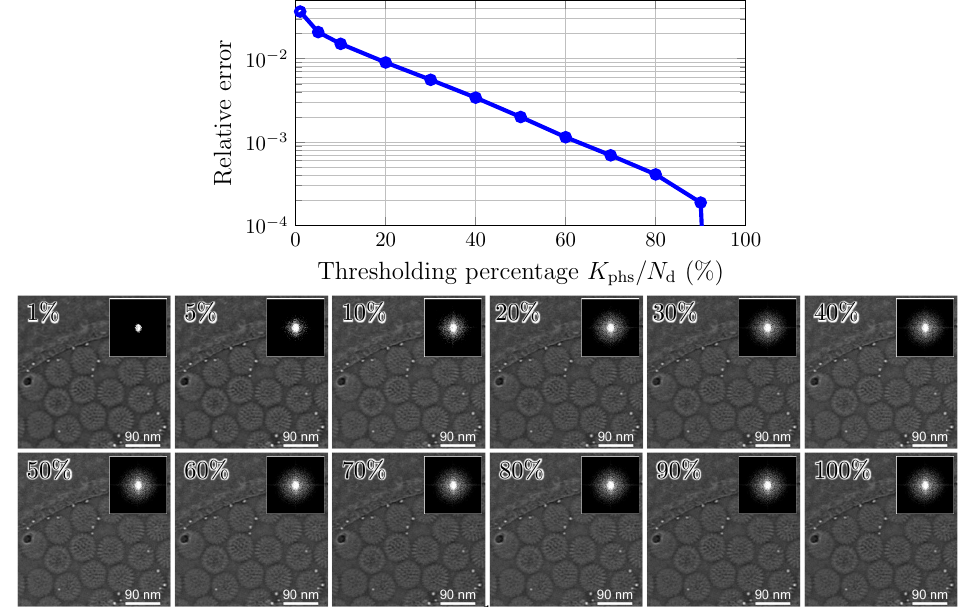}}
    \caption{\textbf{Low-complexity of the reference Rotavirus phase image.} The relative error is computed between the reference phase image shown in Fig.~\ref{fig:reference_data} and its best $K_{\rm phs}$-term approximation in 2-D DCT domain. Keeping only 1\% of the DCT coefficients and mapping the rest to zero results in an approximation of the reference phase image that is almost indistinguishable to human eyes with approximately 3.5\% relative error.}
    \label{fig:low_complexity_reference}
\end{figure}

Fig.~\ref{fig:low_complexity_reference} shows the relative error between the reference phase image $\bs x$ and its $K_{\rm phs}$-best term approximation $\hat{\bs x}$, \ie $\| \bs x - \hat{\bs x}\|_2/\|\bs x\|_2$. In Fig.~\ref{fig:low_complexity_reference}-bottom we observe that discarding up to 99\% of the DCT coefficients of the phase image results in an approximation image with approximately 3.5\% relative error and that is almost indistinguishable to human eyes.

\clearpage
%%%%%% Start of Section 3 %%%%%%
\subsection{Ablation study: impact of regularisation parameters on the performance of LoRePIE}
\label{sec:ablation_metric_downsampling}
In Sec.~\ref{sec:numerical-results} we set the regularisation parameters $K_{\rm phs} = K_{\rm amp} = 0.05 N_{\rm d}$. In this section we justify that parameter selection. Due to time constraints all the reconstructions in this section are obtained using 10 iterations of the LoRePIE algorithm.

We consider down-sampling factor in the range of $\{1,2,3,4,5\}$, which is equivalent to  subsampling ratios $\{1, 1/4, 1/9, 1/16, 1/25, 1/36\}$. For each down-sampling factor, we obtain 121 reconstructions for every pair of the phase and amplitude regularisation parameters, \ie $K_{\rm amp}/N_{\rm d} \in \{5, 10, 20, \cdots, 100\}$ and $K_{\rm phs}/N_{\rm d} \in \{5, 10, 20, \cdots, 100\}$. We compute the quality of every reconstructed phase image compared to the reference phase image shown in Fig.~\ref{fig:reference_data} in terms of the Structural Similarity Index Measure (SSIM) and the Peak Signal-to-Noise Ratio (PSNR). We also compute the Normalised Root Mean Squared Error (NRMSE) between the reference object and the complex-valued reconstructed object data.

The ${\rm PSNR}(\bs u, \bs v)$ in dB between two signals $\bs u, \bs v \in \bb R^{N}$ is defined as
\begin{equation}\label{eq:psnr_definition}
{\rm PSNR} (\bs u, \bs v) \coloneqq 10 \log_{10}\frac{|\max(\bs u) - \min(\bs u)|^2}{N^{-1}\| \bs u - \bs v \|_2^2},
\end{equation}
where $\max(\bs u)$ and $\min(\bs u)$ return, respectively, the maximum and minimum value of $\bs u$. A perfect reconstruction gives ${\rm PSNR} = \infty$ dB.

We refer the reader to \cite{wang2004image} for the definition and details of SSIM metric. A perfect reconstruction gives ${\rm SSIM} = 1$.

The ${\rm NRMSE}(\bs u, \bs v)$ in dB between two signals $\bs u, \bs v \in \bb R^{N}$ is defined \cite{maiden2009improved} as
\begin{equation}\label{eq:nrmse_definition}
{\rm NRMSE} (\bs u, \bs v) \coloneqq -10\log_{10} \frac{\|\bs u - \gamma \bs v\|^2}{\| \bs u \|_2^2},
\end{equation}
where $\gamma \coloneqq \bs v^* \bs u /\|\bs v\|^2$. A perfect reconstruction gives ${\rm NRMSE} = \infty$ dB. The parameter $\gamma$, makes NRMSE invariant to multiplication of a complex-valued scalar $c \in \bb C$, \ie 

\begin{equation*}
{\rm NRMSE} (\bs u, c\bs v) = -10\log_{10} \frac{\|\bs u - \frac{c^*\bs v^* \bs u}{\|\bs c \bs v\|^2}  c \bs v\|^2}{\| \bs u \|_2^2} = {\rm NRMSE} (\bs u, \bs v).
\end{equation*}
Therefore, NRMSE between a signal and its scaled version is infinity, \ie ${\rm NRMSE} (\bs u, c\bs u) = \infty$ dB.

\begin{figure}
    \centering
    \scalebox{0.87}{\includegraphics[width=\columnwidth]{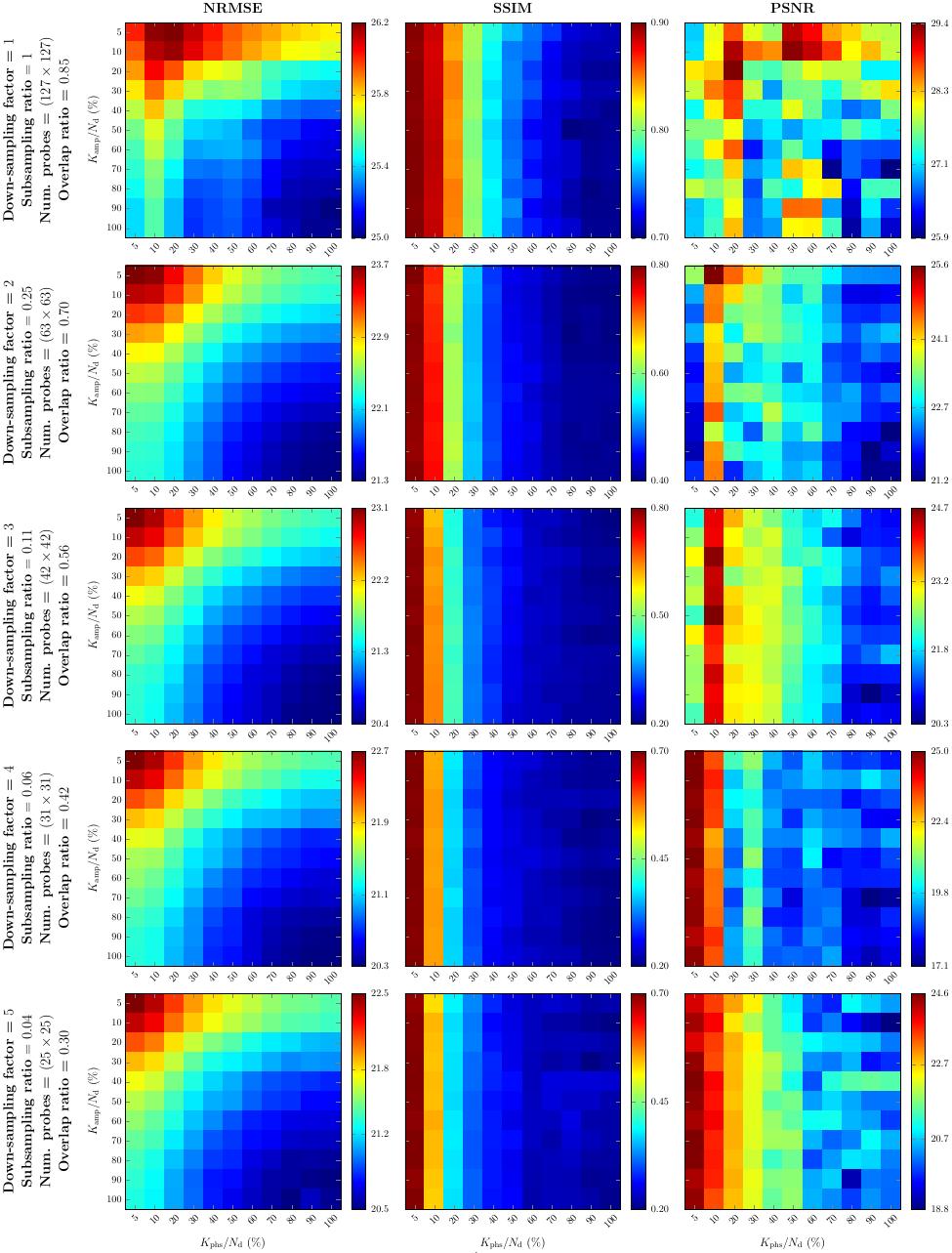}}
    \caption{\textbf{Impact of two regularisation parameters on the quality of phase retrieval for multiple down-sampling factors of probe positions.} LoRePIE with arbitrary regularisation parameters always outperforms ePIE -- \ie when $K_{\rm phs} = K_{\rm amp} = N_{\rm d}$ -- in the sense of NRMSE and SSIM and in the sense of PSNR except for only few outliers. We set $K_{\rm phs} = K_{\rm amp} = 0.05 N_{\rm d}$ for the remaining of the numerical experiments in this paper.}
    \label{fig:ablation_metric_downsampling}
\end{figure}

Fig.~\ref{fig:ablation_metric_downsampling} depicts the quality of the reconstructed data for multiple down-sampling ratios and for the considered range of regularisation parameters. We remind that the NMRSE is computed using the complex-valued object data, while SSIM and PSNR are computed using only the phase component of the object data. We also note that $K_{\rm phs} = K_{\rm amp} = N_{\rm d}$ amounts to the ePIE algorithm, \ie lower-right element in each quality map in Fig.~\ref{fig:ablation_metric_downsampling}. We observe that \textit{(i)} LoRePIE with arbitrary regularisation parameters in the considered range always outperforms ePIE in the sense of the three studied quality metrics with only few exceptions when considering the PSNR metric. \textit{(ii)} when considering the full data, \ie down-sampling ratio = 1, $(K_{\rm phs}/N_{\rm d}, K_{\rm amp}/N_{\rm d}) = (0.2,0.05)$ yields the highest NMRSE = 26.2, while for other down-sampling ratios in $\{2,3,4,5\}$, the pair of $(K_{\rm phs}/N_{\rm d}, K_{\rm amp}/N_{\rm d}) = (0.05,0.05)$ yields the highest NRMSE values. \textit{(iii)} SSIM, which is calculated base on only the phase image, is almost insensitive to the regularisation parameter of the object amplitude. \textit{(iv)} The choice of parameter $K_{\rm phs}/N_{\rm d} = 0.05$ always provides the highest quality phase image in the sense of SSIM quality metric. \textit{(v)} Despite the fact that the quality maps of PSNR are less structured compared to those of SSIM and NRMSE, LoRePIE results in higher values of PSNR, except only few instances that are due to the sensitivity of PSNR to pixel-wise imperfection. \textit{(vi)} Overall, lower regularisation parameters yields higher PSNR values.

Therefore, we find $K_{\rm phs} = K_{\rm amp} = 0.05 N_{\rm d}$ to be, on average, efficient parameter setting for LoRePIE that yields an approximately 2.4 dB, 0.4, and 2.5 dB improvement compared to ePIE in, respectively, NRMSE, SSIM, and PSNR values.

\begin{figure}
    \centering
    \scalebox{1}{\includegraphics[width=\columnwidth]{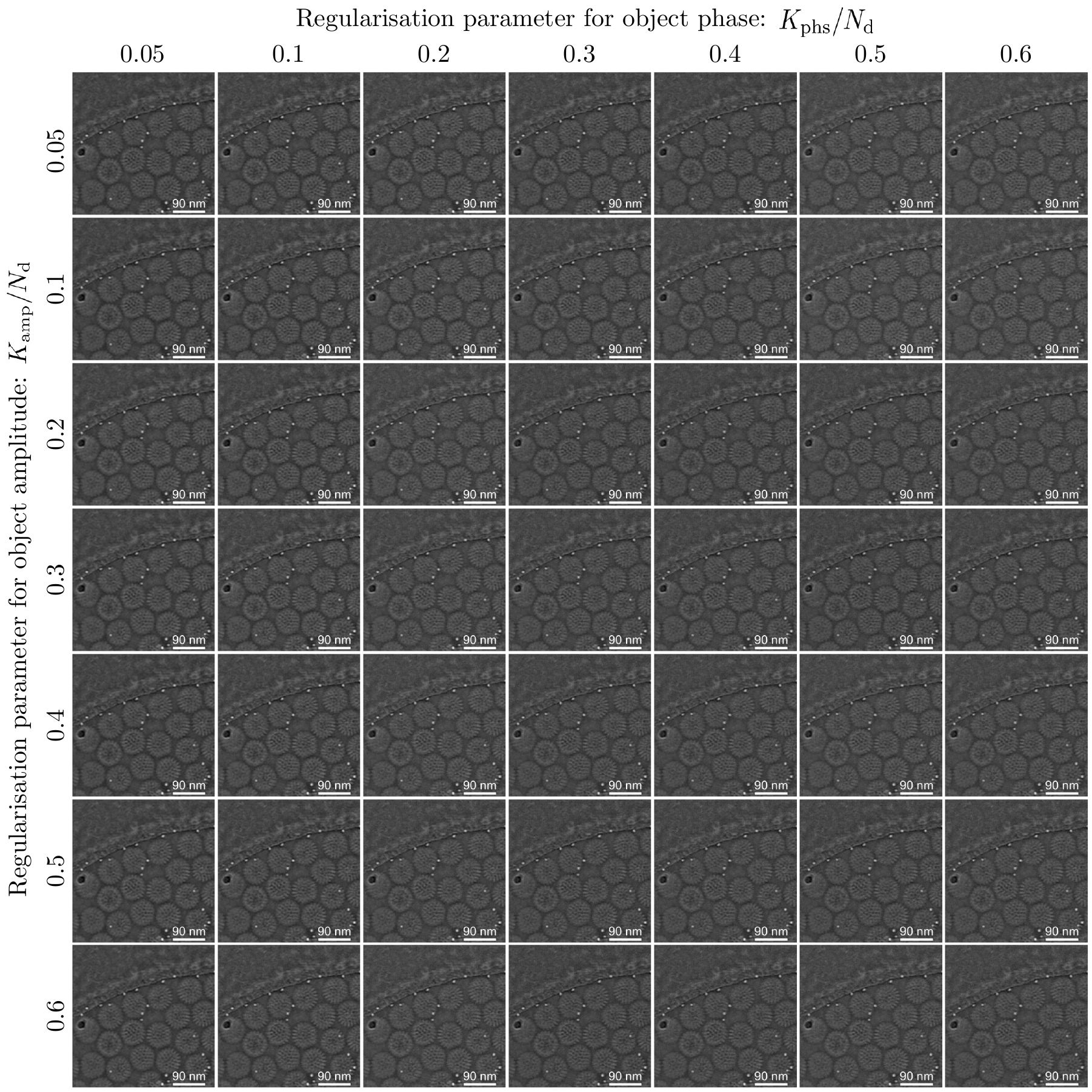}}
    \caption{\textbf{Impact of two regularisation parameters on the quality of reconstructed phase images for down-sampling factor = 1.} This figure is extracted from the first row of Fig.~\ref{fig:ablation_metric_downsampling}.}
    \label{fig:lorepie_object_phase_10itr_downsampling_1}
\end{figure}

\begin{figure}
    \centering
    \scalebox{1}{\includegraphics[width=\columnwidth]{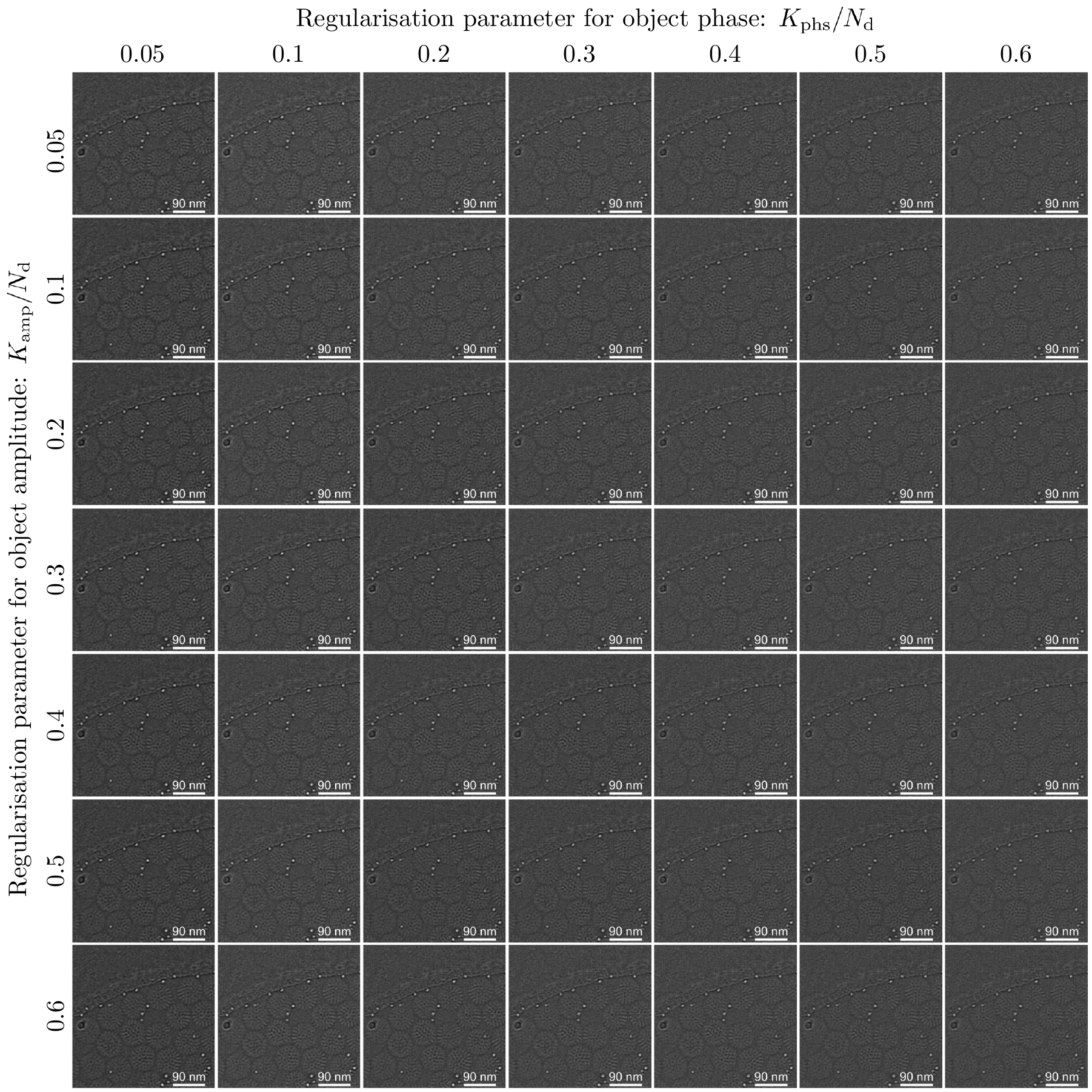}}
    \caption{\textbf{Impact of two regularisation parameters on the quality of reconstructed phase images for down-sampling factor = 2.} This figure is extracted from the second row of Fig.~\ref{fig:ablation_metric_downsampling}.}
    \label{fig:lorepie_object_phase_10itr_downsampling_2}
\end{figure}

\begin{figure}
    \centering
    \scalebox{1}{\includegraphics[width=\columnwidth]{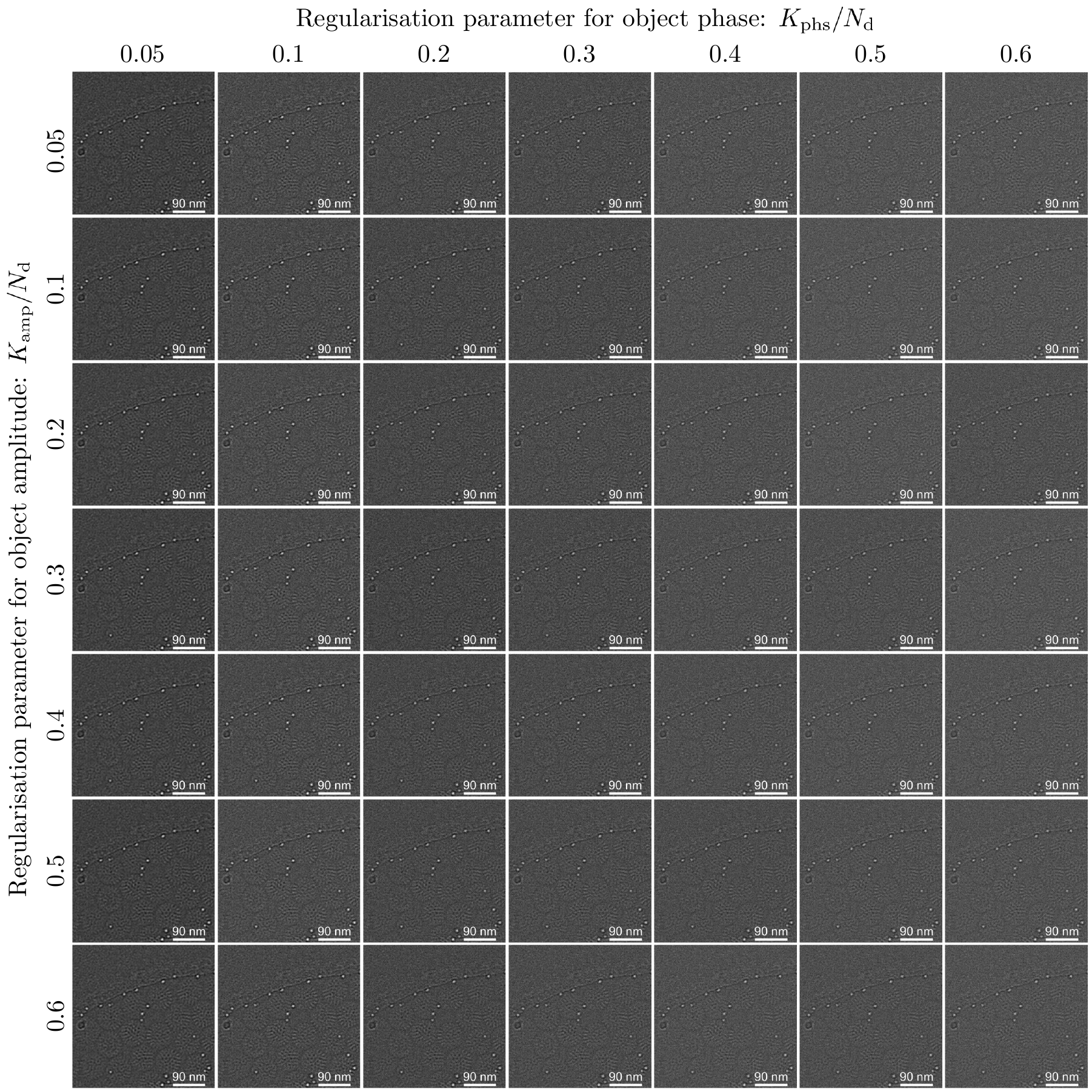}}
    \caption{\textbf{Impact of two regularisation parameters on the quality of reconstructed phase images for down-sampling factor = 3.} This figure is extracted from the third row of Fig.~\ref{fig:ablation_metric_downsampling}.}
    \label{fig:lorepie_object_phase_10itr_downsampling_3}
\end{figure}

\begin{figure}
    \centering
    \scalebox{1}{\includegraphics[width=\columnwidth]{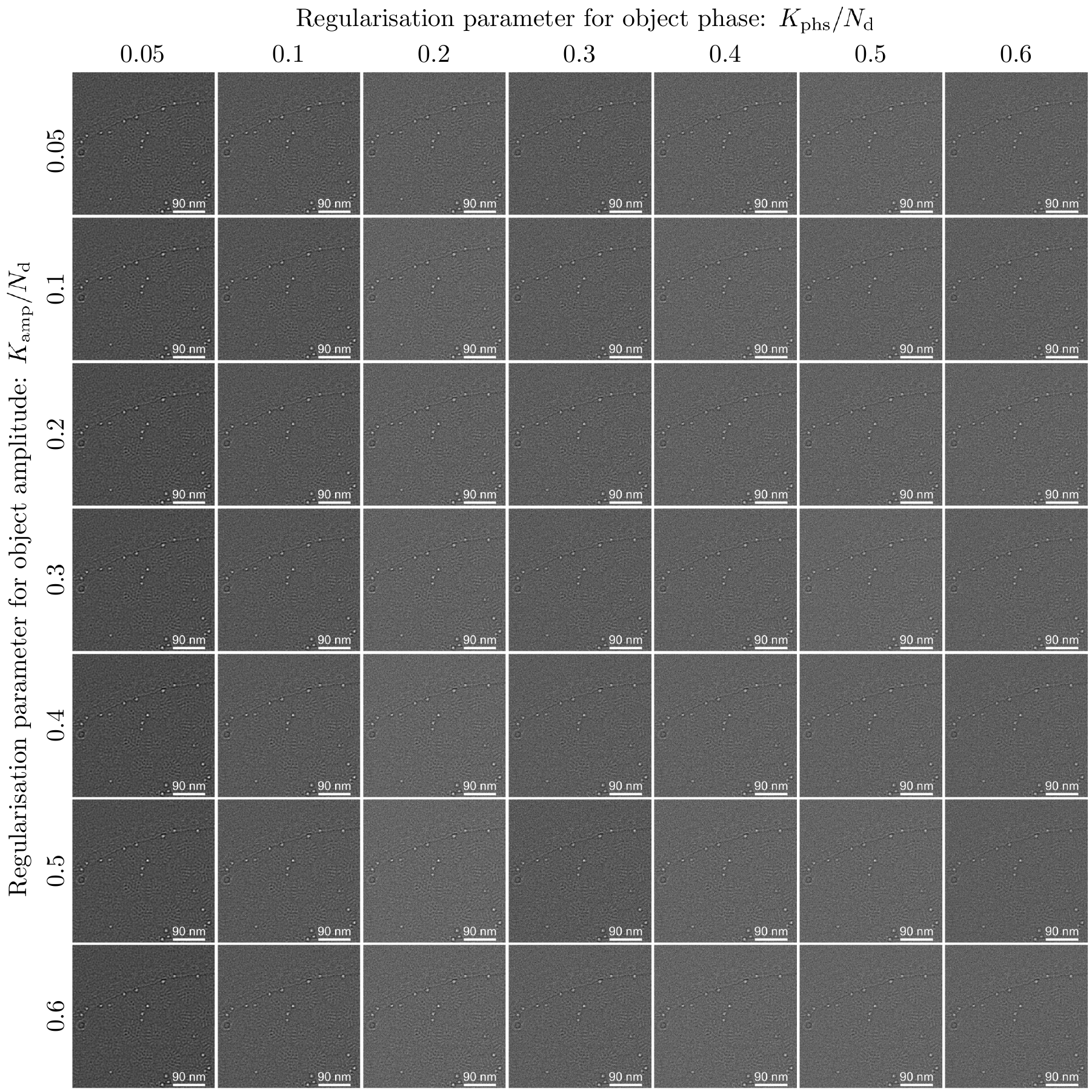}}
    \caption{\textbf{Impact of two regularisation parameters on the quality of reconstructed phase images for down-sampling factor = 4.} This figure is extracted from the fourth row of Fig.~\ref{fig:ablation_metric_downsampling}.}
    \label{fig:lorepie_object_phase_10itr_downsampling_4}
\end{figure}

\begin{figure}
    \centering
    \scalebox{1}{\includegraphics[width=\columnwidth]{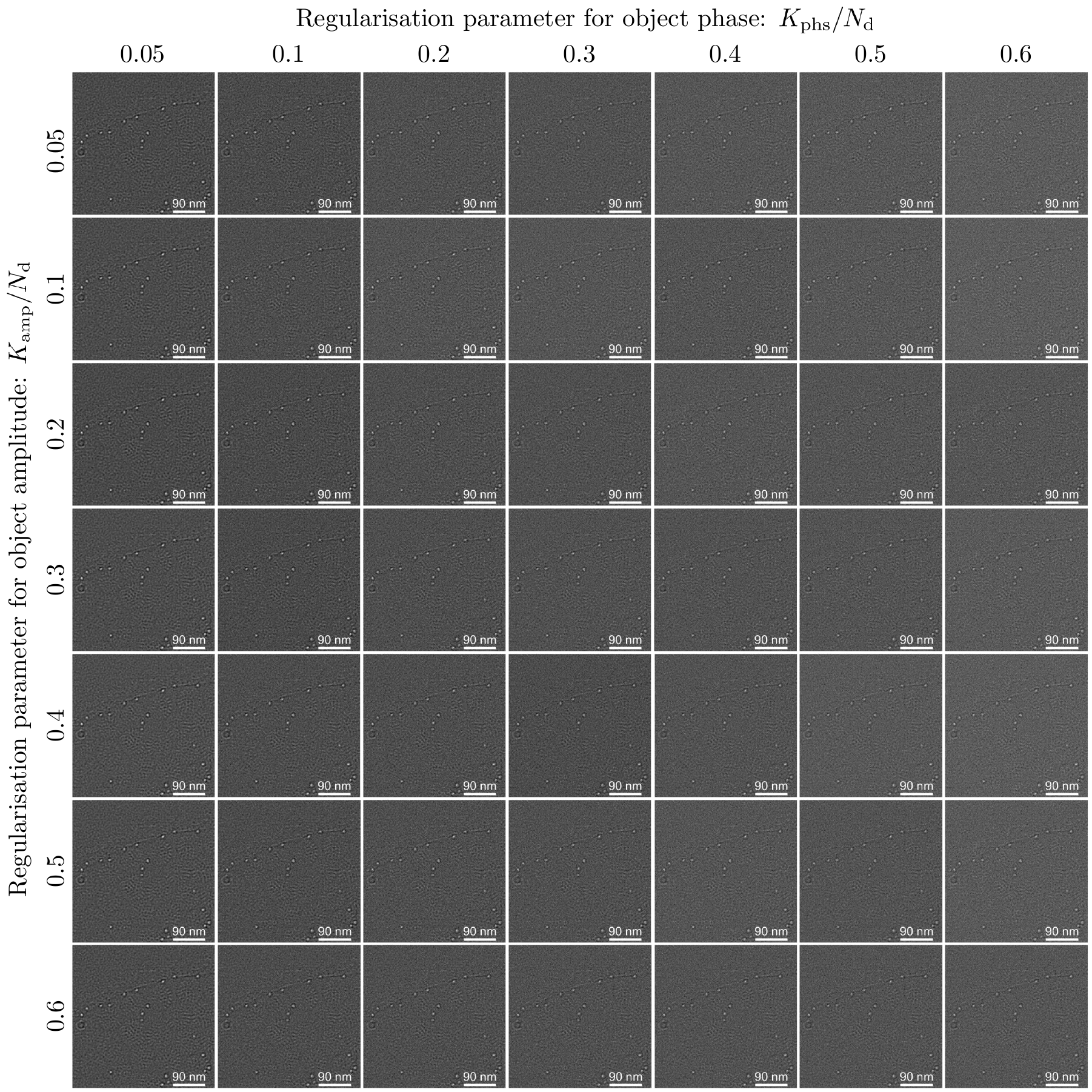}}
    \caption{\textbf{Impact of two regularisation parameters on the quality of reconstructed phase images for down-sampling factor = 5.} This figure is extracted from the fifth row of Fig.~\ref{fig:ablation_metric_downsampling}.}
    \label{fig:lorepie_object_phase_10itr_downsampling_5}
\end{figure}

Furthermore, example reconstructed phase images extracted from Fig~\ref{fig:ablation_metric_downsampling} are shown in Figs.~\ref{fig:lorepie_object_phase_10itr_downsampling_1}, \ref{fig:lorepie_object_phase_10itr_downsampling_2}, \ref{fig:lorepie_object_phase_10itr_downsampling_3}, \ref{fig:lorepie_object_phase_10itr_downsampling_4}, and \ref{fig:lorepie_object_phase_10itr_downsampling_5} for down-sampling factor of, respectively, 1, 2, 3, 4, and 5.

\clearpage
%%%%%% Start of Section 4 %%%%%%
\subsection{Ablation study: impact of number of iterations on the performance of LoRePIE and ePIE}
\label{sec:ablation_kamp_5_kphs_5}
Following the study in Sec.~\ref{sec:ablation_metric_downsampling}, we set $K_{\rm phs} = K_{\rm amp} = 0.05 N_{\rm d}$ and investigate the impact of the number of iterations on the performance of ePIE and LoRePIE algorithms. 

\begin{figure}[tbh]
    \centering
    \scalebox{1}{\includegraphics[width=\columnwidth]{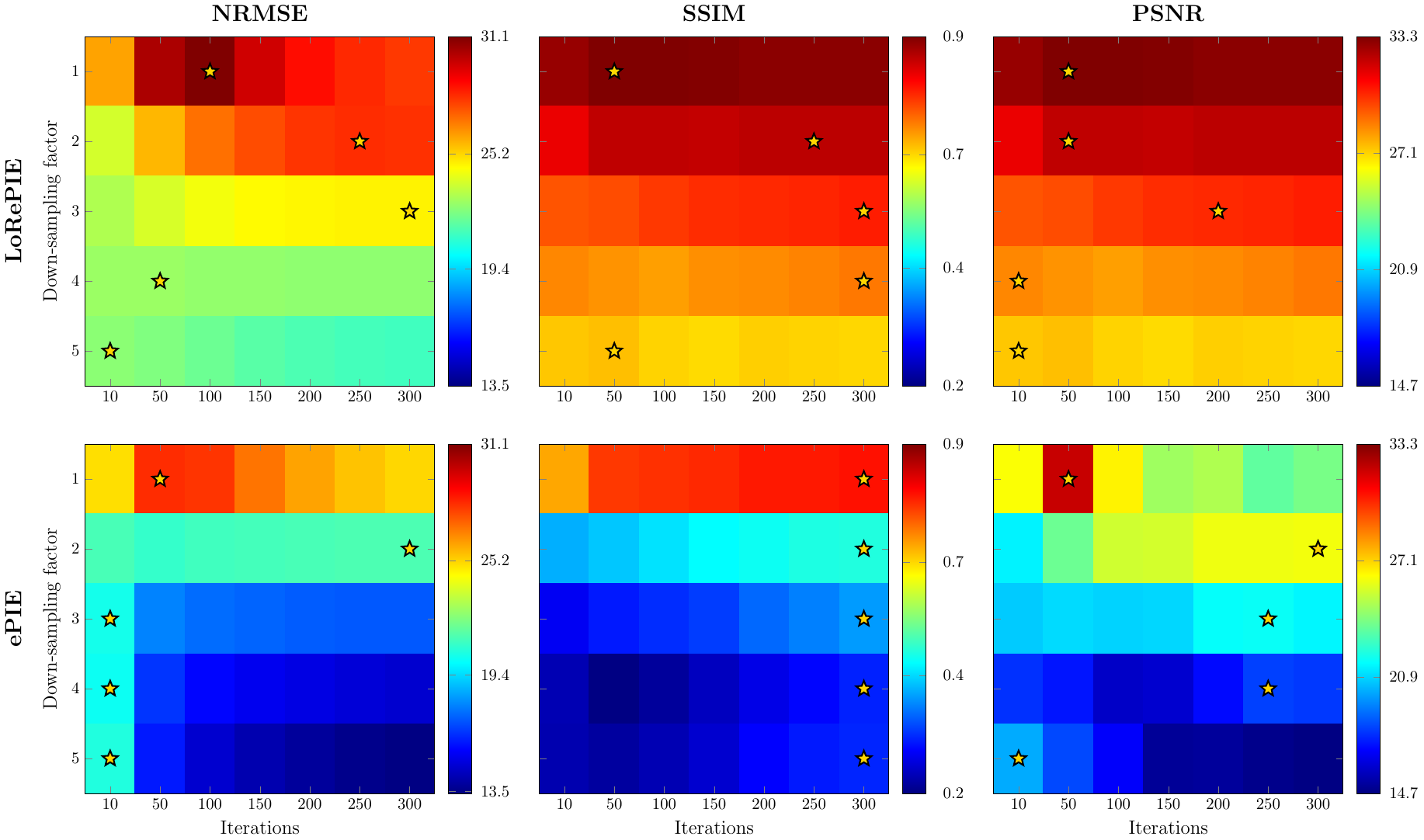}}
    \caption{\textbf{Impact of number of iterations on the quality of reconstructed phase images for various down-sampling factors for ePIE and LoRePIE algorithms.} The highest quality value for each down-sampling factor is indicated by a star. For LoRePIE algorithm we set $K_{\rm phs} = K_{\rm amp} = 0.05 N_{\rm d}$. Examples of reconstructed phase images from this figure are shown in Fig.~\ref{fig:ablation_lorepie_kamp_5_kphs_5_examples} and Fig.~\ref{fig:ablation_epie_examples} for, respectively, LoRePIE and ePIE algorithms.}
    \label{fig:ablation_kamp_5_kphs_5}
\end{figure}

Fig.~\ref{fig:ablation_kamp_5_kphs_5} shows the NRMSE, SSIM, and PSNR values for reconstructed phase images using the ePIE and LoRePIE algorithms, run for various numbers of iterations and down-sampling factors. It is evident that LoRePIE consistently outperforms ePIE, regardless of the quality metric, across all down-sampling factors and number of iterations. The highest quality value for each down-sampling factor is indicated by a star. It is evident that greater number of iterations does not necessarily yield higher quality reconstruction. Moreover, we do not observe an explicit correlation between the NRMSE, SSIM, and PSNR values. We remind that the NRMSE values were computed on complex-valued object data, while the SSIM and PSNR values were computed only on normalised object phase data.

\begin{figure}[tbh]
    \centering
    \scalebox{1}{\includegraphics[width=\columnwidth]{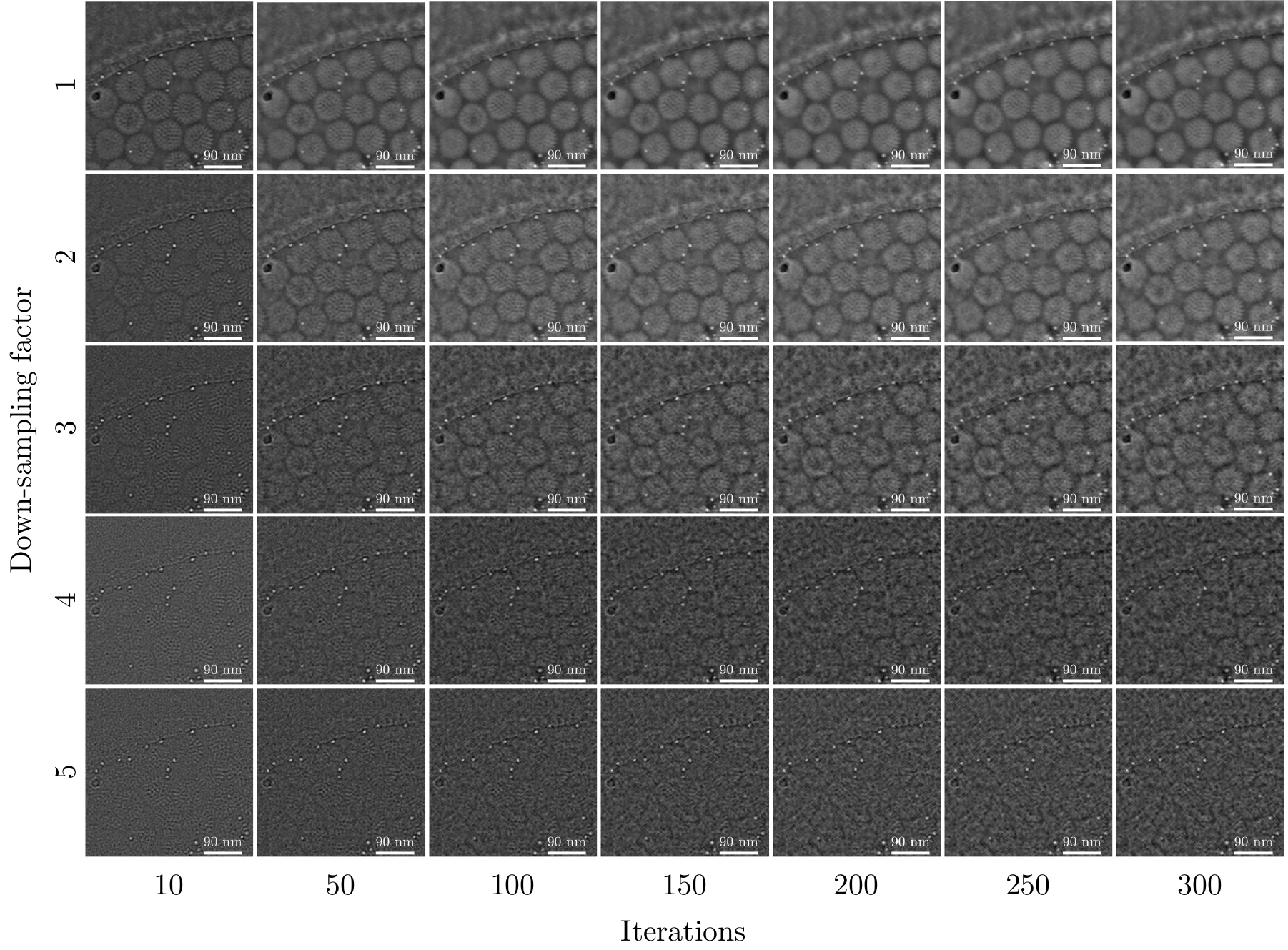}}
    \caption{\textbf{Examples of reconstructed phase images using LoRePIE algorithm for various down-sampling factors with $K_{\rm phs} = K_{\rm amp} = 0.05 N_{\rm d}$.} These examples are extracted from the top row of Fig.~\ref{fig:ablation_kamp_5_kphs_5}. In general, increasing the number of LoRePIE iterations leads to smoother reconstructed phase images. The results suggest that optimal number of LoRePIE iterations depends on the down-sampling factor, and consequently, the overlap ratio.}
    \label{fig:ablation_lorepie_kamp_5_kphs_5_examples}
\end{figure}

\begin{figure}[tbh]
    \centering
    \scalebox{1}{\includegraphics[width=\columnwidth]{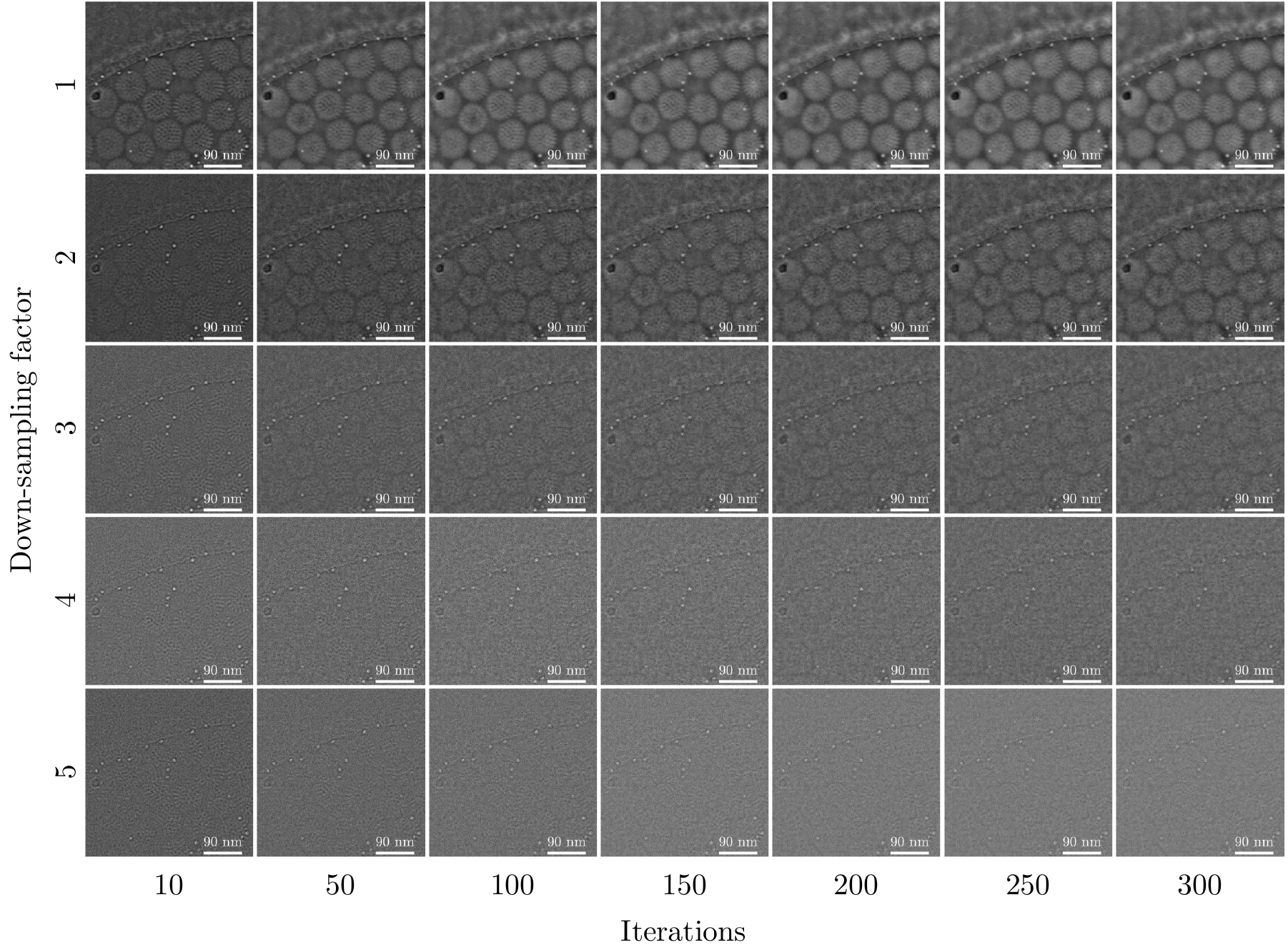}}
    \caption{\textbf{Examples of reconstructed phase images using ePIE algorithm for various down-sampling factors.} These examples are extracted from the bottom row of Fig.~\ref{fig:ablation_kamp_5_kphs_5}. Similarly to LoRePIE algorithm, increasing the number of ePIE iterations leads to smoother reconstructed phase images. The results suggest that optimal number of ePIE iterations depends on the down-sampling factor, and consequently, the overlap ratio.}
    \label{fig:ablation_epie_examples}
\end{figure}

Examples of reconstructed phase images from the top and bottom row of Fig.~\ref{fig:ablation_kamp_5_kphs_5} are shown, respectively, in Fig.~\ref{fig:ablation_lorepie_kamp_5_kphs_5_examples} and Fig.~\ref{fig:ablation_epie_examples}. In general, increasing the number of LoRePIE or ePIE iterations leads to smoother reconstructed phase images. The results suggest that optimal number iterations for both algorithms depends on the down-sampling factor, and in turn, the overlap ratio.

Based on the observations in Figs.~\ref{fig:ablation_kamp_5_kphs_5}, ~\ref{fig:ablation_lorepie_kamp_5_kphs_5_examples}, and ~\ref{fig:ablation_epie_examples}, 50 iterations of the LoRePIE and ePIE algorithms appear to offer a reasonable trade-off between reconstruction quality and time. Therefore, for the presented results in Sec.~\ref{sec:numerical-results} of the main document we ran LoRePIE and ePIE algorithms for 50 iterations.
\vfill
\clearpage

\vfill
\clearpage

\vfill
\clearpage

\vfill
\clearpage
%%%%%% Start of Section 5 %%%%%%
\subsection{Ablation study: impact of update step size parameters on the performance of LoRePIE and ePIE}
\label{sec:ablation_learning_rate_sweep}

Based on our study in Secs.~\ref{sec:ablation_metric_downsampling} and \ref{sec:ablation_kamp_5_kphs_5}, we set $K_{\rm phs} = K_{\rm amp} = 0.05 N_{\rm d}$ and set the number of ePIE and LoRePIE iterations to 50. In this section, we investigate the impact of the update step size parameters for object ($\alpha_{\rm o}$) and probe ($\alpha_{\rm p}$) in ePIE and LoRePIE algorithms.

\begin{figure}[tbh]
    \centering
    \scalebox{1}{\includegraphics[width=\columnwidth]{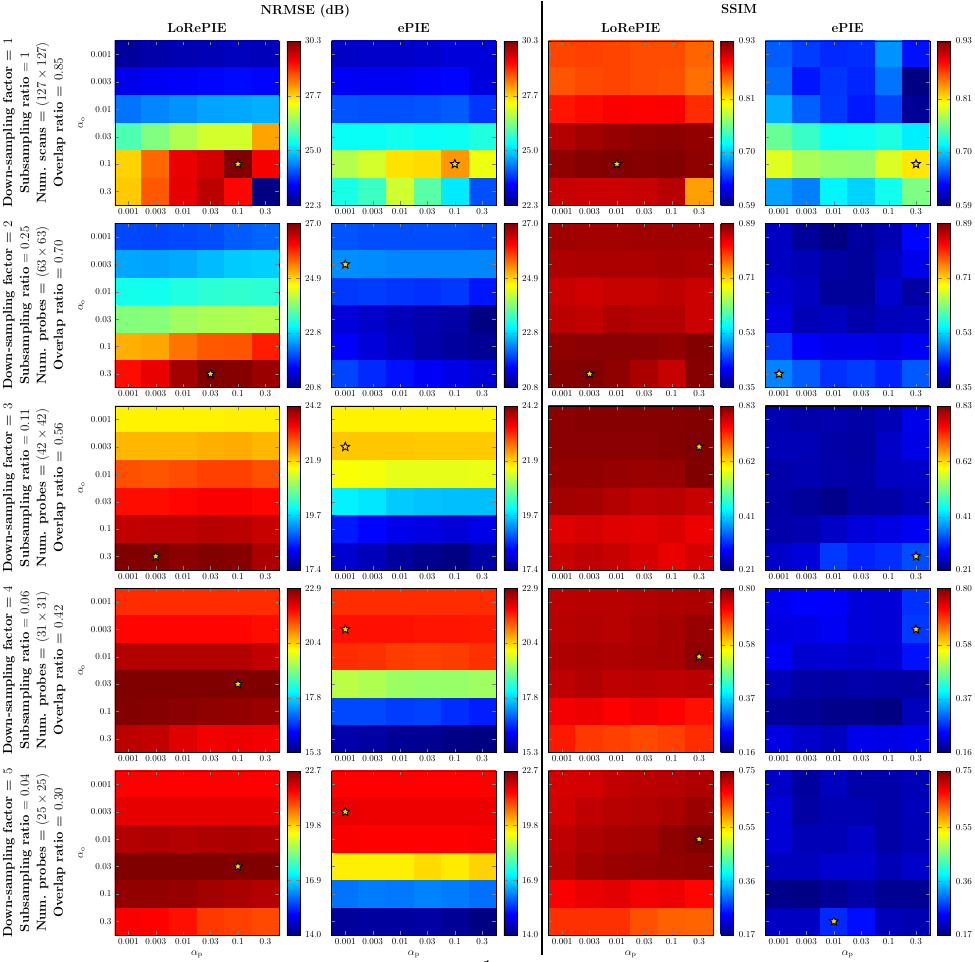}}
    \caption{\textbf{Impact of update step size parameters on the performance of LoRePIE and ePIE algorithms.} The highest quality value for each down-sampling factor is indicated by a star. For LoRePIE algorithm we set $K_{\rm phs} = K_{\rm amp} = 0.05 N_{\rm d}$. Both algorithms are run for 50 iterations. We choose SSIM maps for determining optimal values of update step size parameters and presenting the main results in Sec.~\ref{sec:numerical-results}.}
    \label{fig:ablation_learning_rate_sweep_all}
\end{figure}

The NRMSE and SSIM values for ePIE and LoRePIE reconstructions are plotted in Fig.~\ref{fig:ablation_learning_rate_sweep_all} for various down-sampling factors. LoRePIE consistently outperforms ePIE across the examined range of update step size parameters in terms of both NRMSE and SSIM. As noted earlier in Figs.~\ref{fig:ablation_metric_downsampling} and ~\ref{fig:ablation_kamp_5_kphs_5}, NRMSE and SSIM do not always align when determining the optimal update step size. NRMSE is calculated on the complex-valued object data, while SSIM is based on the normalized phase component of the object. Since the phase is more critical than the amplitude, and visual inspection suggests that the SSIM values better correlate with the visual quality of the object phase, we select the optimal update step size parameters based on the SSIM maps in Fig.~\ref{fig:ablation_learning_rate_sweep_all}. The results presented in Sec.~\ref{sec:numerical-results} of the main document are obtained using the update step size values marked by stars on the right side of Fig.~\ref{fig:ablation_learning_rate_sweep_all}.

\end{document}